\documentclass[preprint,graphicx]{aastex}

\usepackage{emulateapj5}
\usepackage{apjfonts}

\slugcomment{ApJ in press}

\shorttitle{Suprime-Cam Weak Lensing Survey (I)}
\shortauthors{Miyazaki et al.}

\begin{document}
\title{A Subaru Weak Lensing Survey I: Cluster Candidates
and Spectroscopic Verification }
\author{
Satoshi Miyazaki\altaffilmark{1}, 
Takashi Hamana\altaffilmark{1}, 
Richard S. Ellis\altaffilmark{2}, \\
Nobunari Kashikawa\altaffilmark{1}, 
Richard J. Massey\altaffilmark{2},
James Taylor\altaffilmark{3},
Alexandre Refregier\altaffilmark{4}
}
\email{satoshi@subaru.naoj.org}

\altaffiltext{1}{National Astronomical Observatory of Japan, Mitaka, Tokyo 181-8588, Japan}
\altaffiltext{2}{California Institute of Technology, 105-24 Astronomy,
Pasadena CA 91125 USA}
\altaffiltext{3}{University of Waterloo, Department of Physics \& Astronomy,
 Waterloo, Ontario N2L 3G1, Canada}
\altaffiltext{4}{Service d'Astrophysique CEA Saclay, Bat. 709 F-91191 Gif sur
 Yvette, France}
 
\begin{abstract}
We present the results of an ongoing weak lensing survey conducted
with the Subaru telescope whose initial goal is to locate and study the 
distribution of shear-selected structures or {\it halos}. Using a
Suprime-cam imaging survey spanning 21.82 deg$^2$, we present a 
catalog of 100 candidate halos located from lensing convergence 
maps. Our sample is reliably drawn from that subset of our survey area,
(totaling 16.72 deg$^2$) uncontaminated by bright stars and edge
effects and limited at a convergence signal to noise ratio of 3.69. To 
validate the sample detailed spectroscopic measures have been made 
for 26 candidates using the Subaru multi-object spectrograph, FOCAS. 
All are confirmed as clusters of galaxies but two arise as the superposition 
of multiple clusters viewed along the line of sight. Including data available 
in the literature and an ongoing Keck spectroscopic campaign, a total of 
41 halos now have reliable redshifts.  For one of our survey fields, the 
XMM LSS \citep{pierre04} field, we compare our lensing-selected halo 
catalog with its X-ray equivalent. Of 15 halos detected in the XMM-LSS field, 
10 match with published X-ray selected clusters and a further 2 are
newly-detected and spectroscopically confirmed in this work.  Although
three halos have not yet been confirmed, the high success rate within
the XMM-LSS field (12/15) confirms that weak lensing provides a
reliable method for constructing  cluster catalogs, irrespective of
the nature of the constituent galaxies or the intracluster medium.
\end{abstract}


\section{Introduction}
Clusters of galaxies represent the most massive bound systems in
the cosmos. Although they result from non-linear structure evolution, 
the departure from linear growth is modest compared to that for 
less massive objects. As a result, simple analytic models can provide 
an accurate indication of their expected number density at various
redshifts. This is the primary reason why cluster of galaxies are
considered to be valuable cosmological probes.

Cosmological attention has focused on the redshift-dependent number 
of clusters, $N(z)$, whose mass exceeds a certain threshold. This is 
one of the most straightforward observables, and is a function of the
cluster mass function, $d^2n/dMdz$, and the evolution of comoving
volume $dV/d\Omega(z)$. The mass function is obtained from the growth
rate of density fluctuations, $\delta(z)$, numerically
\citep{jenkinsetal01} under the assumption of a particular theory of 
structure formation, e.g. the currently-popular cold dark matter (CDM) model. 
Since both $\delta(z)$ and $dV/d\Omega(z)$ are dependent on
the cosmological model, useful constraints could be estimated by 
comparing $N(z)$ with various model predictions. To make progress, 
e.g. on the dark energy equation of state parameter $w$, 
data on several thousand clusters to $z\simeq$1 is thought to be 
required \citep{levineetal02,wangetal04}, and maintaining an
accurate and uniform mass threshold is critical.

Most early work focused on selecting clusters optically, with detection 
techniques that have improved over the decades: e.g. matched-filter 
\citep{postmanetal96}, red-sequence \citep{gladdersetal00}, cut-and-enhance 
\citep{gotoetal02}. Optically selected samples have traditionally
suffered from uncertainties in the optical richness - mass relation,
although there has been recent progress in calibrating
the closely related richness-velocity dispersion relation using the
large  maxBCG sample of clusters identified in the Sloan Digital Sky
Survey \citep{beckeretal07}.

So far, X-ray samples have been the most popular cosmological probes
e.g. \cite{bohringeretal01}, \cite{ikebeetal02}, \cite{reiprichetal02}. 
Luminosity ($L_X$) or temperature ($T_X$)-limited samples 
offer simpler selection functions because the observables, $L_X$,
$T_X$, are a fair estimate of the cluster mass, calibrated through 
empirical scaling relations. The derived mass does depend, however,
on the assumed dynamical state of the system. Unrelaxed clusters,
arising for example from recent mergers, will introduce scatter in 
the scaling relation. A recent study by \cite{smithetal05} points out 
that at least half of 10 z $\sim$ 0.2 cluster cores show unrelaxed 
features and a scatter of $\sigma \sim 0.4$ around the mean scaling relation.

Weak gravitational lensing, which analyses the coherent shear 
pattern of background galaxies, can potentially provide 
estimates of the cluster mass {\it regardless of its dynamical state
or the properties of the constituent galaxies}. For some years,
the method has been used to calibrate mass obtained from 
X-ray data \citep{smail97}. \cite{allenetal03} concluded that X-ray
mass measurements are consistent with those from weak lensing,
particularly for relaxed systems which thus offer a useful
cosmological probe\citep{allenetal04}. 

A natural extension of this progress is thus to consider selecting
clusters directly from weak lensing signals. The development of
panoramic imaging surveys has now made this a practical
proposition. \cite{wittmanetal01} reported the first discovery of a cluster
located from a weak lensing convergence map, during the
course of conducting the Deep Lens Survey \citep{wittmanetal02}. 
\cite{miyazakietal02a} later undertook a systematic search of mass 
concentrations on a 2 deg$^2$ field using the Suprime-Cam imager
on Subaru. They detected several significant (S/N$>4$) candidates, 
one of which was later spectroscopically identified as a cluster at 
$z$ =0.41. \cite{hetterscheidtetal05} investigated 50 randomly-selected 
VLT FORS1 fields, spanning 0.64 deg$^2$ in total, and found 5 
shear-selected candidates, each associated with an overdensity
in luminosity. The first results from the Deep Lens Survey, based
on an area of 8.6 deg$^2$ have also recently emerged \citep{wittmanetal06}.

The above pioneering studies have demonstrated that clusters can
be located directly via weak lensing. However, key issues, including 
the optimum selection threshold, the rate of spurious detection and 
the degree of mass completeness at a given redshift, crucial for any 
eventual cosmological application, remain unresolved.

At the present time, theoretical studies offer the only insight into
these issues. Projection is one of the most troublesome aspects of a weak
lensing survey, because of the relatively broad window function.
Unrelated structures contributing to the signal would lead to an
overestimate of the cluster mass. Moreover, as the noise in the 
convergence map arises largely from shot noise in the ellipticity
distribution of background galaxies, some fraction of genuine clusters 
might be missed in a shear selected catalog. N-body and ray-tracing 
simulations \citep{whiteetal02,hamanaetal04,hennawiandspergel05}
have concluded that, for systems whose convergence signal lie above a 
4 standard deviation ($\sigma$) threshold, 60-75\% of clusters can
reliably recovered (completeness). Likewise, for peaks detected
in the simulated data using typically-used algorithms, 60-75 \% 
represent genuine clusters (efficiency). The difference in these 
figures between the various studies is largely due to differences in
the lower mass limit adopted in the studies. 

This series of papers is motivated by the need to address these
key issues observationally. The survey we describe is a natural
and ongoing extension of the 2 deg$^2$ survey of \cite{miyazakietal02a};
to date a total field of 21.82 deg$^2$ has been imaged. This first paper 
describes the imaging survey and discusses the validation of the
candidates found, both via spectroscopic verification and comparison
with clusters located via X-ray techniques. Later papers in the series 
will extend the spectroscopic follow-up to the full sample and will 
consider the feasibility of deriving cosmological constraints from 
both this survey and future enhanced versions. 
We note that a similary motivated program has been initiated by 
\cite{maturietal06} ,  \cite{schirmeretal06} and
\cite{dietrichetal07}, all of which made used of the imaging data
taken by 2.5 m VLT survey telescope. We compare the their conclusions
with our own in this paper.

We note that
a similarly motivated program has been initiated by \cite{maturietal06} 
and \citet{schirmeretal06} whose conclusions we compare with our own 
in this paper.

A plan of the paper follows. We discuss the imaging
strategy and data analysis in \S2, including construction of the cluster 
catalog and its reliability. In \S3, we describe our initial spectroscopic
follow-up with Subaru and address the completeness by comparing 
X-ray selected clusters on one of our survey area where relevant
X-ray data is available. We summarize our conclusions in \S4.

\section{Imaging Observations \& Data Analysis}

\subsection{Survey Fields}

\begin{table*}
\caption{Suprime-Cam Weak Lensing Survey Fields. 
\label{tab:survey_field}}
\begin{center}
\begin{tabular}{lllccccccc}
\tableline\tableline\noalign{\smallskip}
Field  & RA  &  DEC & Area\tablenotemark{a}& Secure Area\tablenotemark{b} &  Seeing\tablenotemark{c} & $\rho_{gal}$\tablenotemark{d} & $T_{R}$\tablenotemark{e} & $T_{C}$\tablenotemark{e} & $T_{N}$\tablenotemark{e} \\
       &     &      & deg$^2$ & deg$^2$ & arcsec& arcmin$^2$ & ksec &  ksec & ksec \\ \hline
DEEP02       & 02:30 & 00  & 1.39  & 0.73 & $0.70\pm 0.06$ & $33.5\pm 6.1 $ &
&     &    \\
SXDS     & 02:18 &--05 & 1.12 & 0.83 & $0.68\pm 0.06$ & $47.7 \pm 5.7 $ &    &     & 100 \\
XMM-LSS     & 02:26 &--04 & 2.80 & 2.24 & $0.55\pm 0.07$ & $46.0 \pm 6.7 $  &    &     & 10 \\
Lynx         & 08:49 & +45 & 1.76 & 1.28 & $0.80\pm 0.08$ & $30.7\pm 7.3 $ & 64  & 300 & 140 \\
COSMOS       & 10:02 & +01 & 1.92 & 1.41 & $0.54\pm 0.03$ & $37.1\pm 2.1$ &    &     & 30  \\
Lockman Hole & 10:52 & +57 & 1.85  & 1.57 & $0.60\pm 0.14$ & $39.3\pm 7.8 $ &200 & 300 & 100 \\ 
GD140        & 11:36 & +30 & 1.83  & 1.50 & $0.71\pm 0.17$ & $29.3\pm 12.9 $ & 33 &     &     \\
PG1159-035   & 12:04 &--04 & 1.43  &1.19 & $0.75\pm 0.05$ & $23.4\pm 3.6 $ & 51 &     &     \\
13 hr Field  & 13:34 & +38 & 2.06  &1.72 & $0.74\pm 0.17$ & $29.6\pm 9.6 $ &110 & 120 & 130 \\ 
GTO2deg$^2$  & 16:04 & +43 & 2.01 & 1.53 & $0.67\pm 0.04$ & $38.0\pm 3.6 $ & 26 &     &     \\ 
CM DRA      & 16:34 & +57 & 1.38  & 0.99 & $0.72\pm 0.12$ & $28.4\pm 8.4 $ & 47 &     &     \\
DEEP16       & 16:52 & +36 & 1.20 & 0.93 & $0.76\pm 0.08$ & $26.4\pm 4.0 $ &    &     &     \\
DEEP23       & 23:30 & 00  & 1.07 & 0.80 & $0.58\pm 0.01$ & $36.3\pm 1.3 $ &    &     &     \\
\noalign{\smallskip}\tableline\noalign{\smallskip}
Total        &       &     & 21.82 & 16.72   & &    &  &     &     \\
\end{tabular}
\tablenotetext{a} {Field area covered by several Suprime-Cam pointings.} 
\tablenotetext{b} {Secure area used for halo sample (see text)}
\tablenotetext{c,d}{Seeing (FWHM) and galaxy densities $\rho_{gal}$ 
refer to average values of the constituent pointings. The scatter is listed 
in terms of a standard deviation.}
\tablenotetext{e}{ $T_{R}$, $T_{C}$, and $T_{N}$ represent X-ray exposure
times of ROSAT, Chandra and XMM-Newton, respectively.} 
\end{center}
\end{table*}

In order to evaluate the efficiency of our weak lensing survey for locating
cluster halos, we considered that a comparison with a sample of X-ray 
selected clusters would be highly advantageous \citep{henry00}.
Therefore, our survey fields were primarily selected to contain X-ray data
as shown in  Table~\ref{tab:survey_field}.  

We set a minimum ROSAT exposure time, $T_R$, of 25 ksec ensuring a 
detection limit of $L_x(0.5-2.0 keV) \sim 2 \times 10^{43}$erg/s at 
$z\simeq$0.5 (for $H_0=75$,$\Omega_M=0.3$, $\Omega_{\Lambda}=0.7$). 
This corresponds to $M \sim 10^{14}M_{\odot}$, which is well matched to 
the likely mass detection limit of our weak lensing survey \citep{miyazakietal02a}.
More sensitive X-ray missions, XMM-Newton and Chandra, have been
surveying ROSAT fields, in part, to deeper limits. These include the  ``Lynx'',  
``Lockman Hole'' and ``UK 13 hr deep field'' in our target list. XMM is also 
actively involved in international campaigns of multi-wavelength wide field 
($> 1 deg^2$) observations such as the ``COSMOS'' , ``XMM-LSS'' and
``SXDS'' fields which we also included. Among these, the XMM-LSS field
\citep{pierre04} is of particular interest given its panoramic area and
published cluster catalogs \citep{valtchanovetal04,willisetal05,pierreetal06}. 

Finally, we included the DEIMOS DEEP2 survey fields \citep{davisetal03}
where spectra of $\sim$ 50,000 faint galaxies will eventually become
available. This will enable close correlations of lensing mass and various
measures of the luminosity density as was recently pioneered for the
COSMOS field \citep{massey07}. Although X-ray data is not available 
for these DEEP fields at the current time, such observations are
planned and will likely become available soon.

\subsection{Observations}

\vspace{0.3cm}
\centerline{{\vbox{\epsfxsize=8.5cm\epsfbox{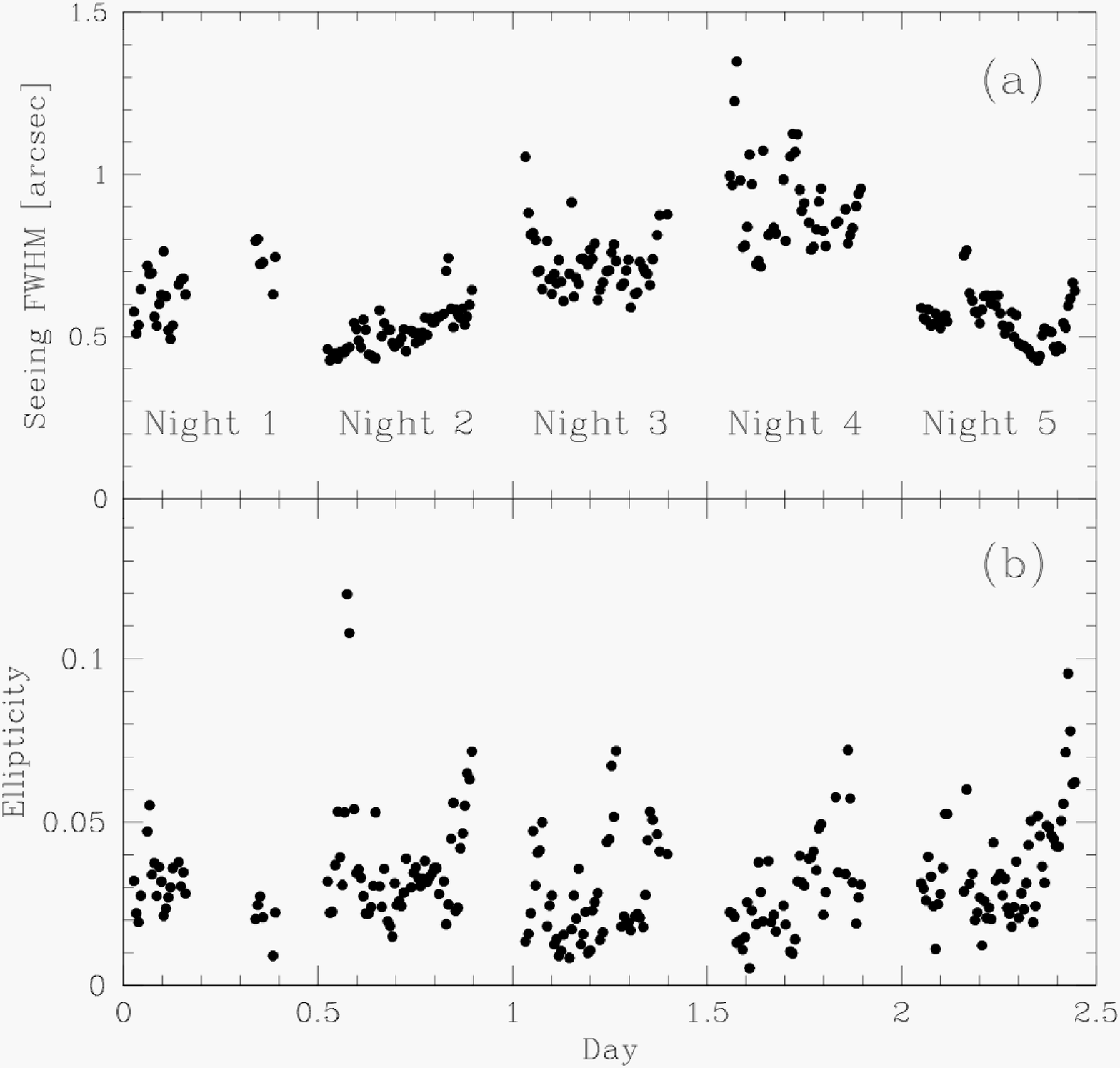}}}}
\figcaption{
Seeing (a) and raw stellar ellpticities (b) measured during the
observing run of a intensive program (May 1-4 and September 27, 2003). 
Average values derived from 700$\sim$1000 stars over the entire field
are plotted. The abscissa shows the observing time (in days) with the
portion of each day shifted arbitrarily for clarity. The median seeing
is 0.65 $\pm$ 0.14 arcsec, which is typical for the survey.
\label{fig:fwhmee}}
\vspace{0.3cm}

The imaging observations were largely carried out on May 1-4 and September 27,
2003 as part of an ``Intensive Program'' of Subaru Telescope. The Lynx field
was observed on January 29-30, 2003, SXDS data was obtained from the public archive
and the COSMOS field observed on February 18 and 21, 2004. 

The Suprime-Cam field size is 0.25 deg$^2$ and all observations were 
conducted with the $R_c$ filter (except the COSMOS field which was observed 
in $i'$-band to enable direct comparison with ACS$/$HST's F814W images).
The total exposure time was 30 minutes for each pointing, taken 
via four 7.5 minute exposures in a dither pattern of spacing $\sim$ 1 arcmin. 

Fig.~\ref{fig:fwhmee} (top) shows the seeing (FWHM) of each 7.5 minute
exposure over a typical five night observing sequence. The bottom panel
shows ellipticities of moderately bright unsaturated stars which provide a
measure of the PSF anisotropy; the raw ellipticity is mostly 2 - 4 \% 
(and occasionally $>$ 5 \%). We discuss the question of the PSF anisotropy 
in Appendix~\ref{sec:imagequality}. Thus far, we have surveyed over 
21.82 deg$^2$ as summarized in Table~\ref{tab:survey_field}.

The useful field area excludes the surroundings of bright stars and galaxies 
and field boundaries. Individual pointings whose seeing was worse than 
0.9 arcsec were also excluded; this occurred for only 5 \% of the clear time
(see section~\ref{sec:wlana}).

\subsection{Data Reduction}

The data reduction procedures for the present survey closely followed
those described in Section 3.1 of \cite{miyazakietal02b}, enhanced
as discussed below.

Normally, with Suprime-Cam images, each CCD exposure is ``mosaic-stacked''  
to yield a single image of a particular pointing. A simple geometrical model is 
used for the focal plane astrometry whose parameters include the effects 
of camera distortion,  the displacement and rotation of each detector from 
a defined fiducial  location and the offset and the rotation of the dithered 
exposures. The best fit parameters are obtained by minimizing the positional 
difference of control stars (70$\sim$100 stars per CCD) held common for 
each exposure. The residual alignment error in this procedure is $\sim$ 0.5 
pixel rms (0.1 arcsec rms).

Such a residual is sufficiently small for most imaging applications.
However, in seeing better than 0.6 arcsec (FWHM), a misalignment of 
0.1 arcsec between two images introduces a $\simeq$2 \% ellipticity on 
the stacked image which is a serious issue for weak lensing studies.

A further improvement is thus necessary. The residual $(\Delta x, \Delta y)$ 
from a reference frame is parameterized as a polynomial function of 
field position $\vec{x} =(x, y)$ as:

\begin{equation}
\Delta x = \sum^{6}_{l=0}\sum^{l}_{m=0} a_{lm,e} x^{l-m}y^m \;\;\;\;
\Delta y = \sum^{6}_{l=0}\sum^{l}_{m=0} b_{lm,e} x^{l-m}y^m
\end{equation}

Pixel values are estimated via linear interpolation of neighboring four 
pixels. The coefficients $a_{lm,e}$ and $b_{lm,e}$ are then obtained by
minimizing the variance of the residuals. A sixth order polynomial
is usually sufficient for this purpose. Each individual CCD image 
is `warped' using  this polynomial correction prior to stacking. This
process reduces the alignment error to $\sim$ 0.07 pixel (0.014 arcsec)
and is similar to the ``Jelly CCD'' model described in \cite{kaiseretal99}. 

Although an external stellar catalog would ideally be used for accurate
astrometry, no such data is available at the relevant faint limits 
(R $>$ 22). We therefore employ the first exposure, corrected by the
simple geometrical model discussed above, as the basic reference
frame.  

We noticed that this refined procedure still introduces some artificial 
deformation. Fig.~\ref{fig:warp}(a) represents the raw image of one CCD 
whereas Fig.~\ref{fig:warp}(b) is that slightly rotated by $7\times 10^{-4}$ 
rad, a typical value, using the mapping described above. Clearly some
degree of artificial deformation is introduced. After some experimentation, 
we found this deformation arises from the undersampled nature
of the Suprime-Cam images.  By adopting 2$\times$2 oversampling prior to
rotation, the resulting ellipticity field shows no sign of image deformation.
However, this is a computationally a very time-consuming solution.

\vspace{0.3cm}
\centerline{{\vbox{\epsfxsize=8.5cm\epsfbox{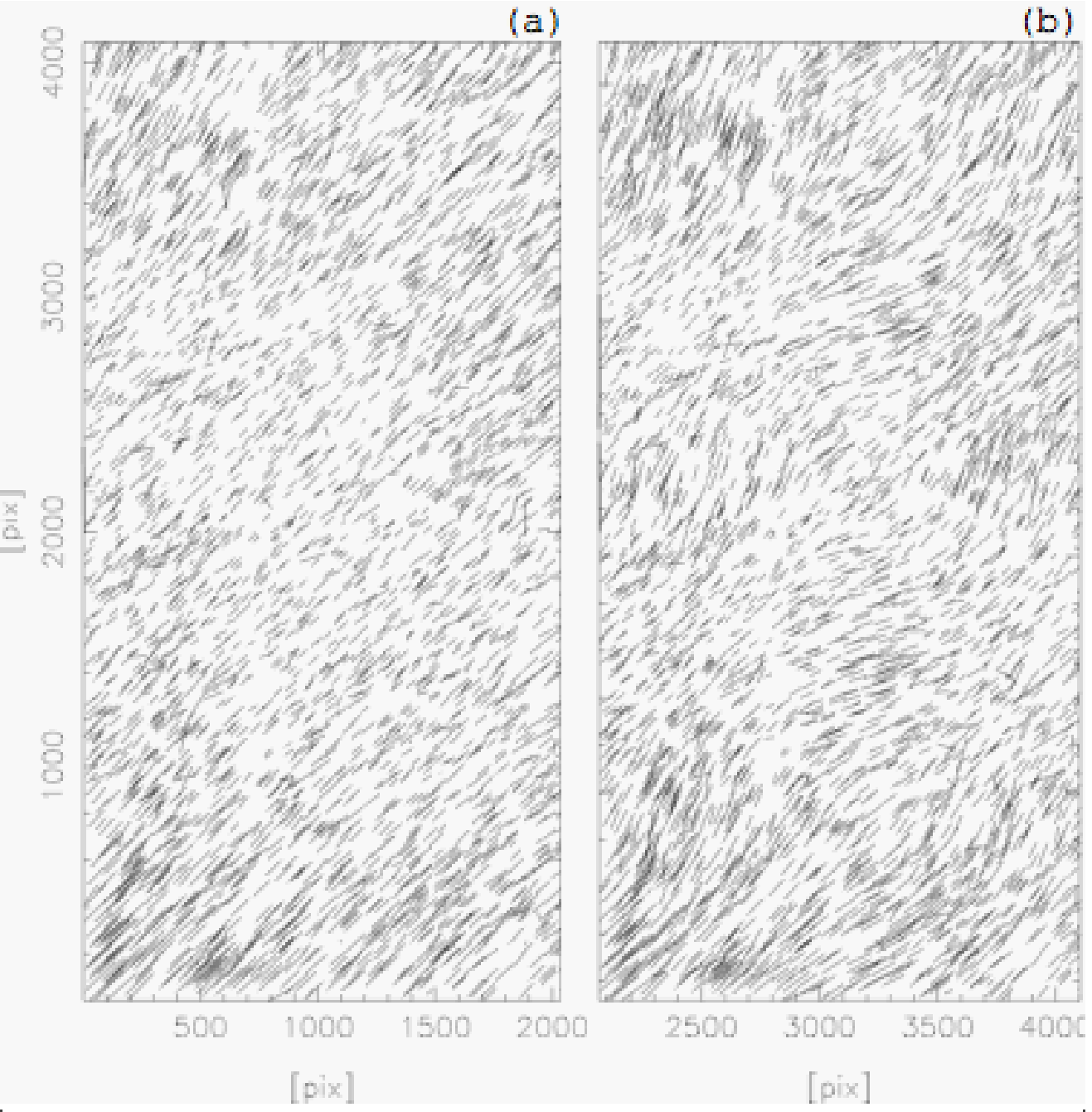}}}}
\figcaption{
Effect of the under-sampled warp correction in an image taken with 
0.7 arcsec seeing: (a) no operation, (b) after rotation by $6.8\times 10^{-4}$ 
rad. A 1.8\% residual is reduced to 0.75 \%.
\label{fig:warp}}
\vspace{0.3cm}

Accordingly, in our final analysis, instead of using oversampled images, we 
modified  the mapping procedure itself. We estimate the  pixel values of the 
target images from 3rd order bi-linear polynomial interpolation of 
4$\times$4 source pixels \footnote{In the actual implementation, we 
employ the Numerical Recipes code {\tt polin2} \citep{pressetal93}. }
rather than the linear interpolation four neighboring pixels in the
previous procedure. This mapping process avoids introducing image 
deformation and is significantly quicker computationally.

\subsection{Galaxy Catalogs} \label{sec:galaxycatalogs}

Object finding and shape measurement was executed on the mosaic-stacked
images using the {\it imcat} software suite developed by Nick Kaiser.  A threshold 
{\it nu}=10 was adopted. Photometric calibration used Landolt standard stars
\citep{landolt92} and the faint standards of \citep{majewskietal94}.
We adopt the Vega magnitude system in the following. 

Galaxies are distinguished from stars via their half light radius, $r_h$, viz:
 
$$r_h > r_{h}^{*} + \sigma_{r_{h}^{*}}$$ 

where $r_{h}^{*}$ and $\sigma_{r_{h}^{*}}$ are the half light radius of 
a stellar image and its rms respectively. The galaxy size distribution is 
shown in  Fig.~\ref{fig:size_gd140}. 

Fig.~\ref{fig:ncnt_gd140} shows the cumulative number density 
of galaxies as a function of  $R_c$-band magnitude. The surface density 
exceeds 50 arcmin$^{-2}$ when the seeing is superb (0.47 arcsec)
and is $\sim15$ arcmin$^{-2}$ in those poor seeing images ($>$0.9 arcsec) 
discarded from our analysis (Fig.~\ref{fig:seeing_gdens_pgamma}(a)). 
Table~\ref{tab:survey_field}  lists the seeing and the galaxy density for 
each field. 

Finally, we masked all objects close to bright stars (within 18 arcsec for 
$b_{USNO-A} < 15$, 90 arcsec for $b_{USNO-A} < 11.7$). Light halos
around bright stars can introduce spurious galaxies.

\vspace{0.3cm}
\centerline{{\vbox{\epsfxsize=8.5cm\epsfbox{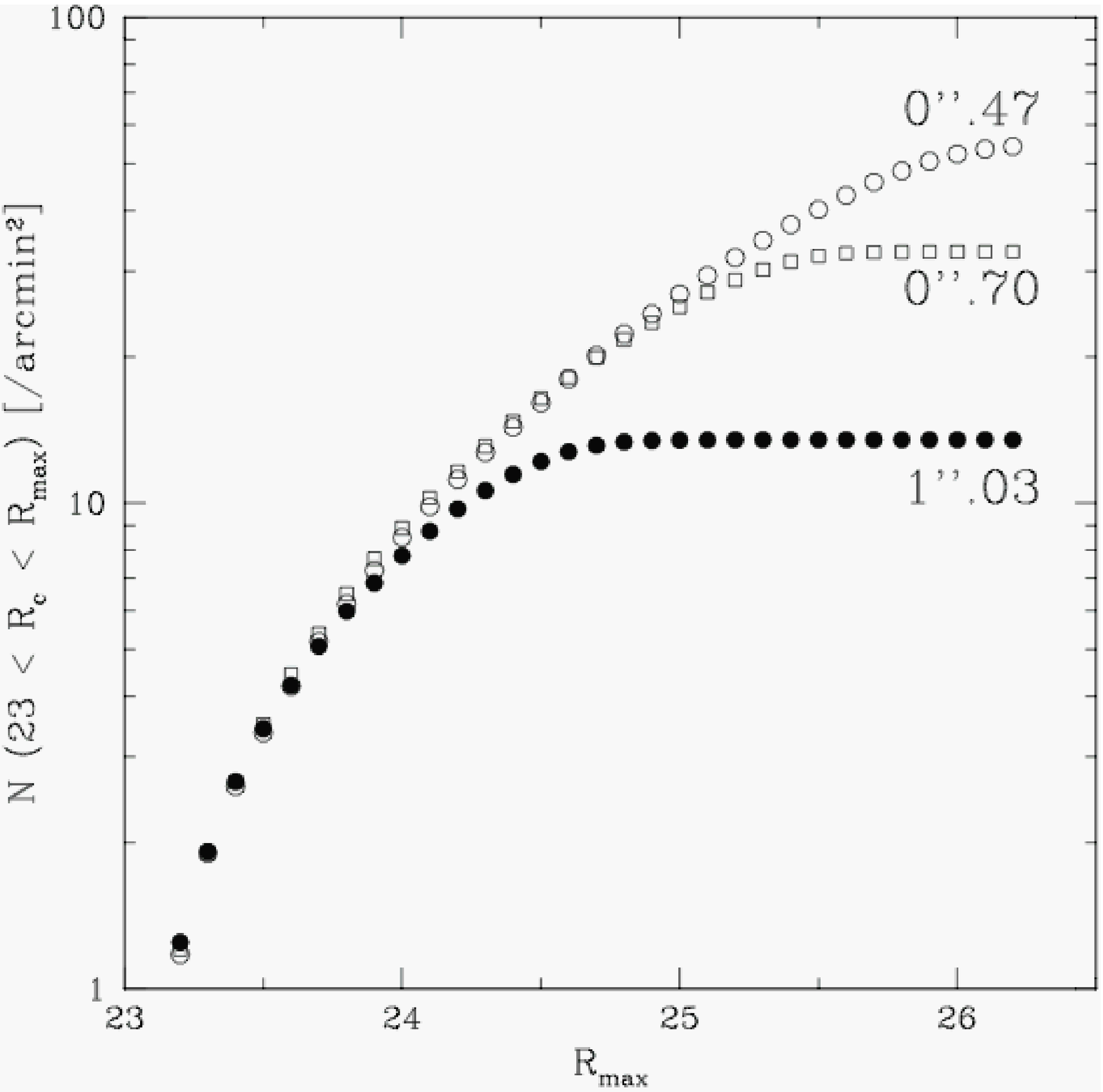}}}}
\figcaption{Cumulative galaxy number counts used in the weak lensing
  analysis. Three representative cases in the GD140 field are 
shown to demonstrate how seeing affects the surface density.
\label{fig:ncnt_gd140}}
\vspace{0.3cm}

\vspace{0.3cm}
\centerline{{\vbox{\epsfxsize=8.5cm\epsfbox{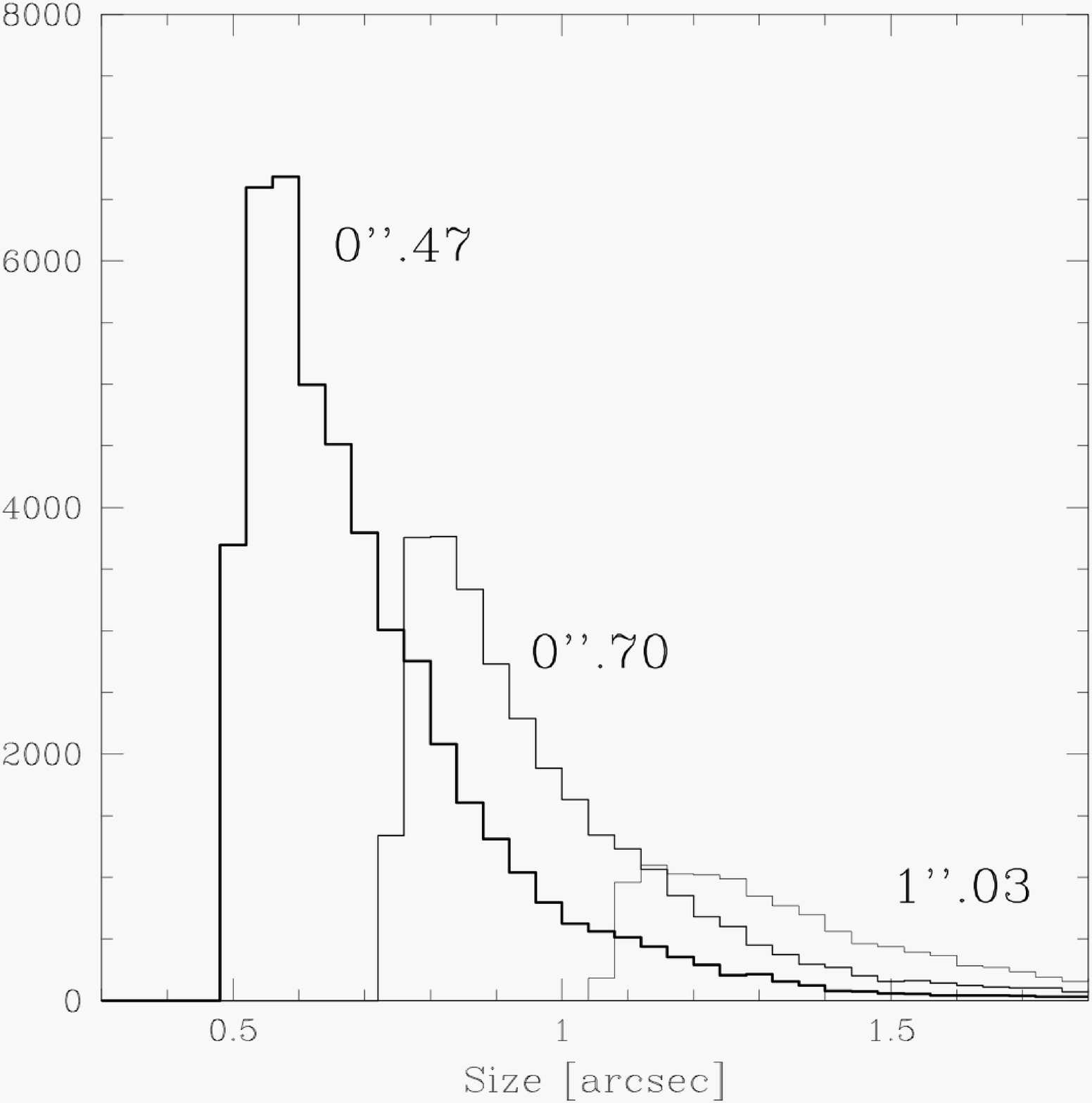}}}}
\figcaption{The size distribution of faint galaxies for the magnitude
  range $23 < R_c <  26 $ observed under under three seeing
  condition. The size is estimated by circular gaussian FWHM here. 
  Those whose size are larger than seeing size are adopted in the
  galaxy catalogs. 
\label{fig:size_gd140}}
\vspace{0.3cm}

\vspace{0.3cm}
\centerline{{\vbox{\epsfxsize=8.5cm\epsfbox{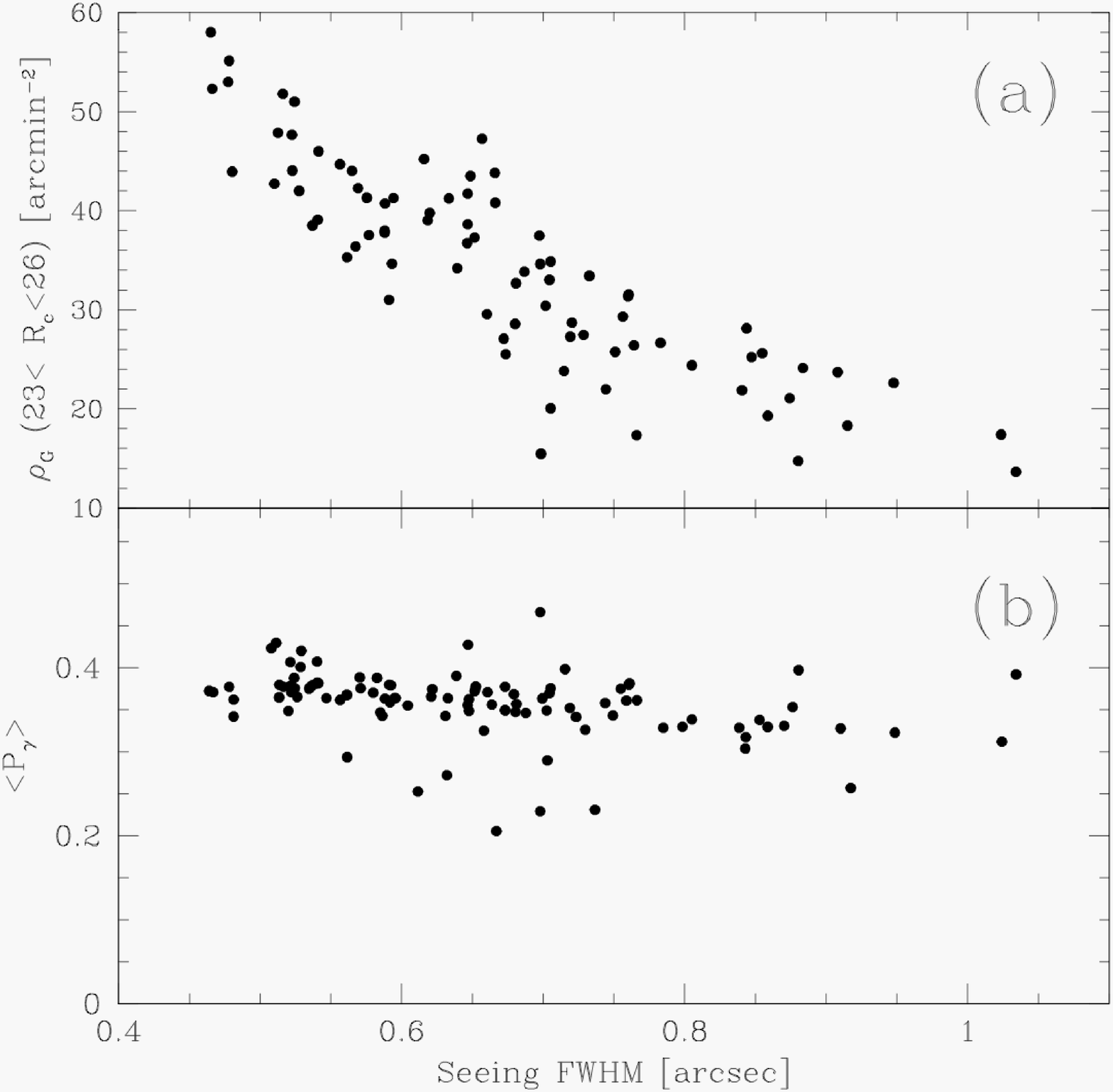}}}}
\figcaption{(a) Seeing dependence of faint galaxy (23 $< R_c <$ 26) 
surface density for all survey pointings except those in the COSMOS field.
(b) Pre-seeing shear polarisability tensor, $P_{\gamma}$, averaged over
the galaxies on each pointing versus the seeing. The seeing dependence is 
satisfactorily small for the selection threshold adopted.
\label{fig:seeing_gdens_pgamma}}
\vspace{0.3cm}

\subsection{Weak Lensing Analysis} \label{sec:wlana}

\subsubsection{Shape Measurements}

Object shapes are represented by the ellipticities, $\vec{e}
= (e_1, e_2)  =  \{I_{11} - I_{22}, 2I_{12}\} / (I_{11}+I_{22})$
where  $I_{ij}$
are Gaussian-weighted quadrupole moments of the surface brightness 
distribution. The point spread function (PSF) of the images is usually
smeared by various instrumental effects such as optical aberrations
and the tracking error of the telescope. The PSF anisotropy is
estimated based on images of stars, and the galaxy images are
corrected so that images of neighboring stars are
re-circularized. Galaxy ellipticities are then corrected as:

\begin{equation}
\label{Psmcorrection}
\vec{e}' =  \vec{e} - \frac{P_{sm}}{P_{sm}^{*}}\vec{e}^{*},
\end{equation}

where the asterisk designates a stellar value, $P_{sm}$ is the smear polarisability
tensor and is mostly diagonal \citep{ksb95}. $(P_{sm}^{-1}\vec{e})^{*}$ is 
evaluated using stars in the field of view and modeled as 5th order bi-polynomial 
function of position. Eqn.\ref{Psmcorrection} then applies this for the galaxy images.
This correction is carried out independently on each pointing. We
further justify the correction procedure in Appendix \ref{sec:imagequality}.

\subsubsection{Shear Estimate}

The shear induced by gravitational lensing, $\vec{\gamma}$, is
diluted by atmospheric seeing. \cite{lk97} developed a prescription to 
convert the observed ellipticities to a 'pre-seeing shear' as

\begin{equation}
\label{pgammacorrection}
\vec{\gamma} =  (P_{\gamma})^{-1}\vec{e}' 
\end{equation}

where $P_{\gamma}$ is the pre-seeing shear polarisability tensor
defined as

\begin{equation}
\label{pgamma}
P_{\gamma}  = P_{sh} - P_{sm} (P_{sm}^{*})^{-1}P_{sh}^{*} .
\end{equation}

$P_{sh}$ is the shear polarisability tensor defined in \cite{ksb95},
and $P_{sh}^{*}$ is the stellar shear polarisability tensor. 

Note that (the inverse of) $P_{\gamma}$ represents the degree of
dilution. Since the $P_{sh}$ and $P_{sm}$ are mostly diagonal, we
replace the tensors in Eqn.~\ref{pgamma} with their trace and evaluate
$P_{\gamma}$ as a scalar. The average value, $<P_{\gamma}>$,
over all galaxies ($23 < R_c < 26 $) of each pointing is shown in
Fig~\ref{fig:seeing_gdens_pgamma}(b).  $<P_{\gamma}>$
decreases slightly as the seeing worsens but the change is
not very large (0.4 to 0.3). This is because we only select larger
galaxies compared with the seeing size. Thus, the dilution factor
is 30$\sim$40 \% regardless of the seeing. In the mean time,
we compare the first component of galaxy ellipticities, $e_1$,
of Suprime-Cam and ACS/HST images taken in the COSMOS field,
and the result is shown in Fig~\ref{fig:suprime_acs}.
The ellipticities  observed by Suprime-Cam are in fact diluted by
36 \% compared with those of ACS, which is consistent with
Fig~\ref{fig:seeing_gdens_pgamma}(b).

\vspace{0.3cm}
\centerline{{\vbox{\epsfxsize=8.5cm\epsfbox{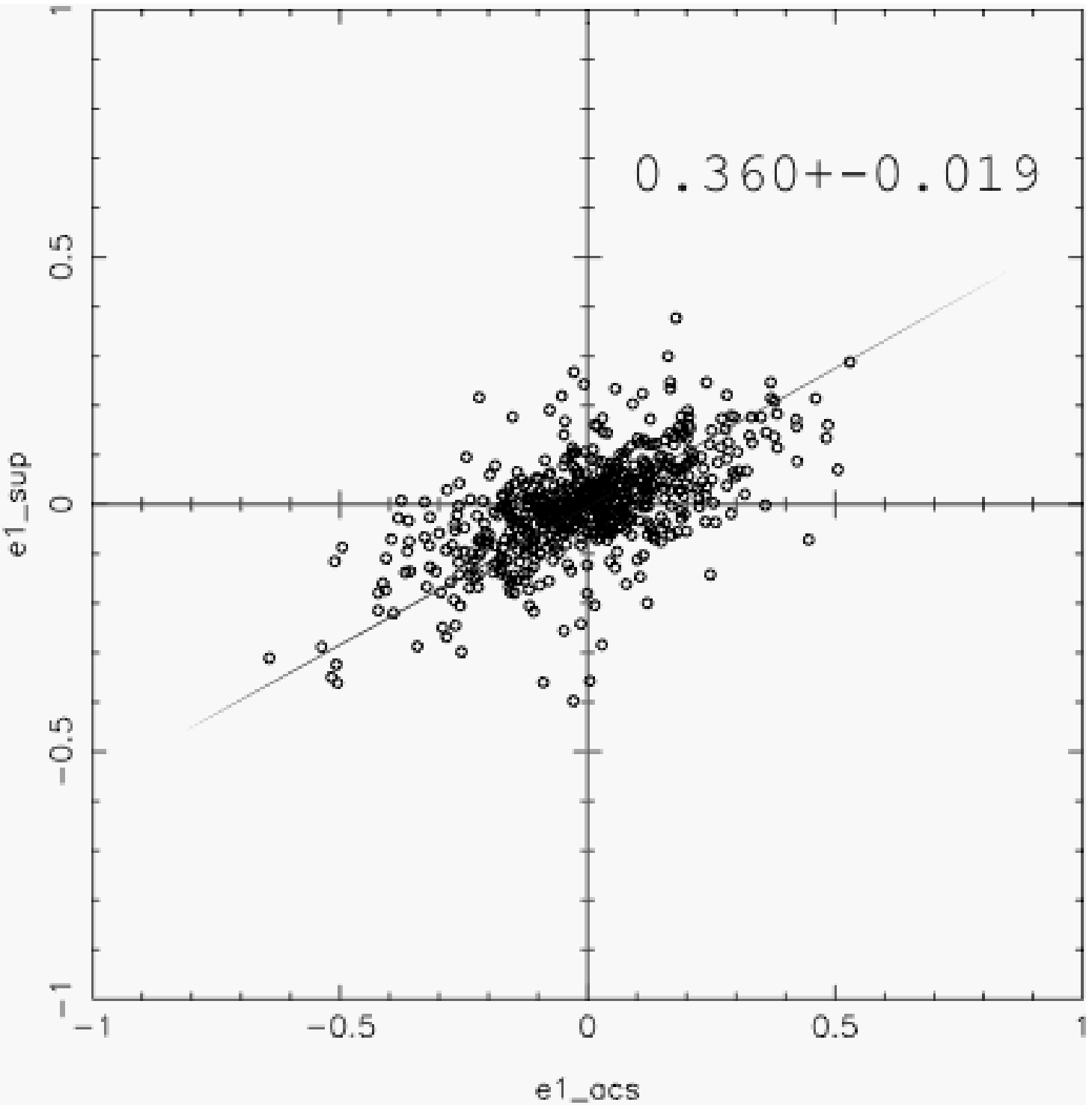}}}}
\figcaption{
Comparison of the first component of ellipticity, $e_1$,  of 
galaxies detected both on HST/ACS (F814W, 34 minutes) and
Suprime-Cam (i'-band, 20 minutes, 0.54 arcsec seeing) images 
in part of the COSMOS field (10 arcmin$^2$). The best fit slope 
and the error are also shown.
\label{fig:suprime_acs}}
\vspace{0.3cm}

To calculate $P_{\gamma}$ as a function of position we employed a
``smoothing''  scheme \citep{waerbeke00,erbenetal01,hamanaetal03}. 
We took the median $P_{\gamma}$ for 20 neighboring galaxies on the 
$r_g$-magnitude plane (where $r_g$ is a measure of object size adopted 
in the {\it imcat} suite). In deriving the mean, the weight $w$ on an
individual measure is taken to be:

\begin{equation}
\label{weight}
w = \frac{1}{\sigma_{\gamma}^2 + \alpha^2},
\end{equation}

where $\sigma_{\gamma}$ is the variance of the raw $\gamma$  
of those 20 neighbors, obtained using the raw $P_{\gamma}$. $\alpha$ 
is the variance of all of the galaxies in the catalog ($\sim$ 0.4). 
In general, the weighted value of a quantity $<A>$ is calculated as 
$<A> = \Sigma_{i=1}^N w_i A_i/\Sigma_{i=1}^Nw_i$ . 

The method we adopt is based on that adopted by \citet{ksb95}. More 
sophisticated methods have since been developed and the variants
are summarized by \citet{heymansetal06}. In their notation, our method 
is very similar to the procedure termed ``LV''.

Based on the results of  the STEP simulation study, \cite{heymansetal06} 
concluded that both the ``KSB+'' method, modified by \cite{hoekstraetal98} 
and implementations of ``BJ02'' method \citep{bernsteinandjarvis02} are 
able to reconstruct input shears to a few percent level. To calibrate
our method, we analyzed the simulated data provided by \cite{heymansetal06}. 
For the model designated ``PSF3'', we underestimate the input shear by
5\% for $\gamma_{true}$ = 0.1 and 0.05, whereas for $\gamma_{true} \le 0.01$, 
the difference, $\gamma_{obs} - \gamma_{true}$  is insignificant. An error 
of 5 \%  in the recovered shear is competitive with most of the methods 
discussed by \cite{heymansetal06} ($\sim$ 7 \% is a typical error). A 5\% 
shear error would induce a similar uncertainty in the mass estimate
of a typical halo. Such an error is considered adequate for the applications
envisaged.

\subsubsection{Kappa Map} 

The dimensionless surface mass density, $\kappa(\vec{\theta})$, is
estimated from the shear field $\gamma(\vec{\theta})$ by the
\cite{kaiserandsquires93} inversion algorithm as:

\begin{equation}
\label{eqn:ks93}
\kappa(\vec{\theta}) = \frac{1}{\pi} \int d^{2}\theta' Re[D^{*}(\vec{\theta} -
  \vec{\theta'})\gamma(\vec{\theta '})],
\end{equation}

where $D(\vec{\theta}) $ is defined as

\begin{equation}
\label{eqn:complexkernel}
D(\vec{\theta}) = \frac{-1}{(\theta_1 - i \theta_2)^2},
\end{equation}

which is a complex convolution kernel for $\kappa$ to obtain
the shear $\gamma$. In the actual implementation, we
smoothed to avoid the effect of noise. A smoothing scale of 
$\theta_G = 1$ arcmin was chosen following the discussion
given by \citet{hamanaetal03}. 

We adopt 15$\times$15 arcsec$^2$ square grids and then
calculate $\kappa(\vec{\theta})$ on each grid using Eqn.\ref{eqn:ks93}
to obtain the $\kappa$ map. In order to estimate the noise of the
$\kappa$ field,  we randomized the orientations of the galaxies in the
catalog and created a $\kappa_{\rm noise}$ map. We repeated this
randomization 100 times and computed the rms value at each grid
point  where $\kappa$ is computed.
Assuming the $\kappa$ error distribution is Gaussian, this rms
represents the 1-$\sigma$ noise level, and thus the measured signal
divided by the rms gives the signal/noise ratio, $\nu(\theta)$, of the
$\kappa$ map at that point.

The Gaussian-smoothed signal/noise map is then searched for mass
concentrations.
\cite{hennawiandspergel05} concluded that a 'truncated' NFW filter applied 
to the aperture mass map, $M_{ap}$, is the most efficient detection
technique. Such optimization may be necessary to improve the efficiency 
of future very wide field surveys where thousands of clusters are sought. 
Here we adopt a simpler technique in order to evaluate its effectiveness
in a direct comparison of lensing and X-ray techniques.

\subsection{Halo Catalog}

\begin{figure*}
\includegraphics[height=17.078cm]{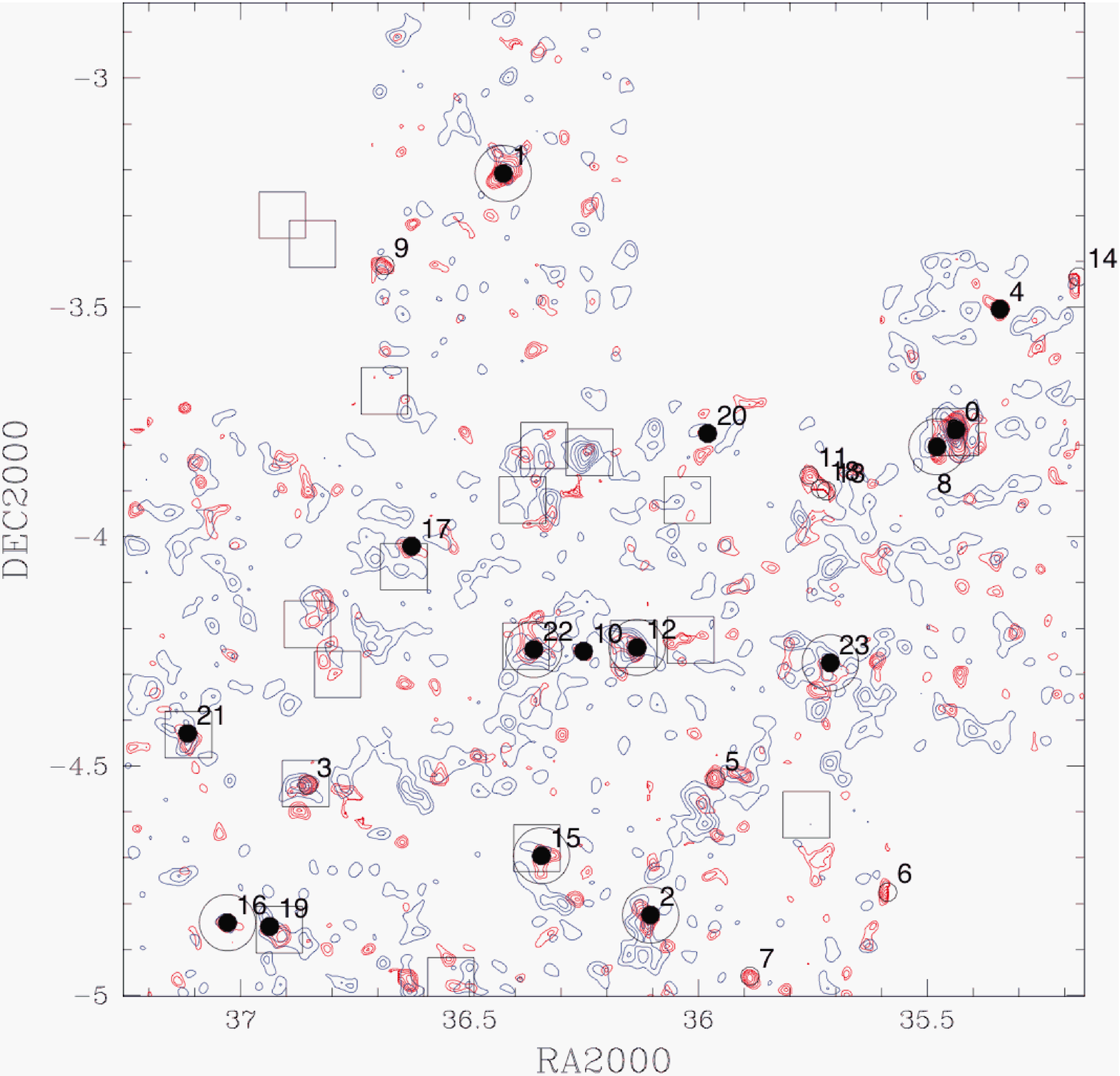}
\caption{The $\kappa$-S/N map for the XMM-wide field (thick red contour). The
lowest contour is set at S/N = 2 with the increments of 0.5. The
smoothing scale, $\theta_G = 1$ arcmin. The thin blue contours show 
the surface number density of visible (21$<R_c<$23) galaxies. Positions 
of detected halos (S/N $>=$ 3.693) are designated by small circles. 
Solid circles represent {\it secure halos} (see text) whereas unfilled 
circles represent samples detected outside the secure area. Squares 
show the location of X-ray selected clusters of galaxies published in 
\cite{valtchanovetal04}, \cite{willisetal05} and \cite{pierreetal06}.
({\bf VWP} data). Large open circles show halos followed up by 
FOCAS and LRIS.
\label{fig:saclay_halomap}}
\end{figure*}

\begin{figure*}
\begin{center}
\centerline{{\vbox{\epsfysize=4.151cm\epsfbox{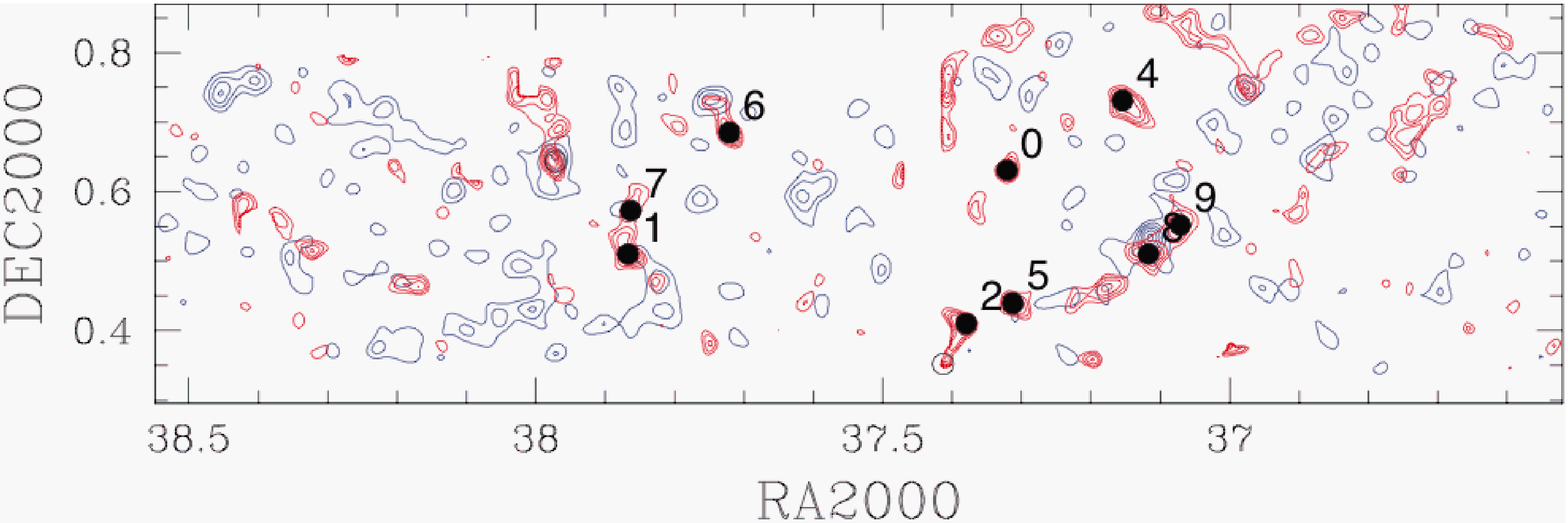}}}}
\figcaption{
DEEP02 field $\kappa$-S/N map (red contour) and 
surface number density of moderately bright (21$<R_c<$23) galaxies
(blue contour). Legends are the same as Fig~\ref{fig:saclay_halomap}. 
\label{fig:deep02_halomap}}
\end{center}
\end{figure*}

\begin{figure*}
\begin{center}
\centerline{{\vbox{\epsfysize=9.296cm\epsfbox{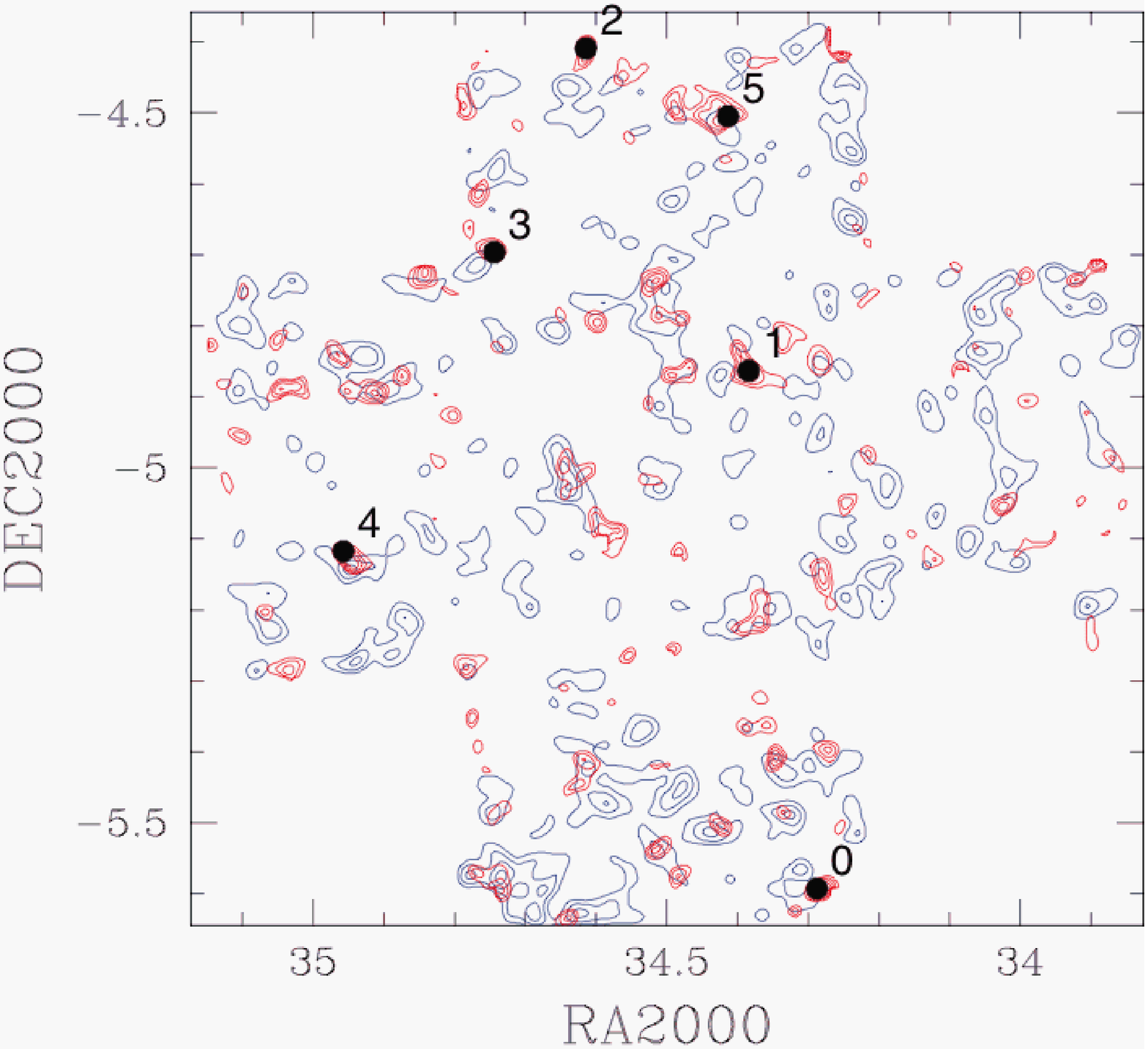}}}}
\figcaption{
SXDS field $\kappa$-S/N map and surface density of galaxies.
(blue contour). 
\label{fig:sxds_halomap}}
\end{center}
\end{figure*}

\begin{figure*}
\begin{center}
\centerline{{\vbox{\epsfysize=11.00cm\epsfbox{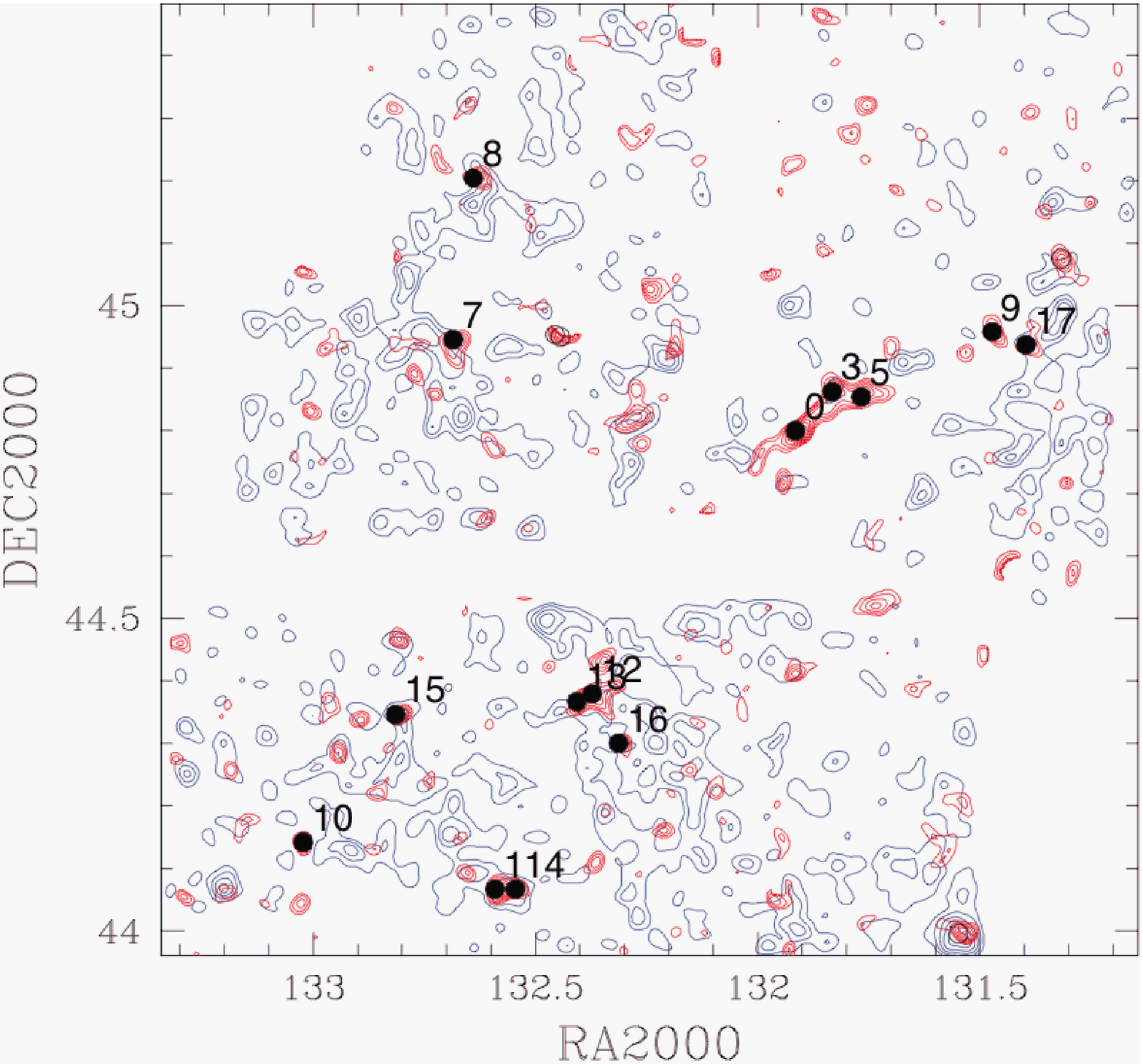}}}}
\figcaption{
Lynx field $\kappa$-S/N map and surface density of galaxies.
\label{fig:lynx_halomap}}
\end{center}
\end{figure*}

\begin{figure*}
\begin{center}
\centerline{{\vbox{\epsfysize=10.507cm\epsfbox{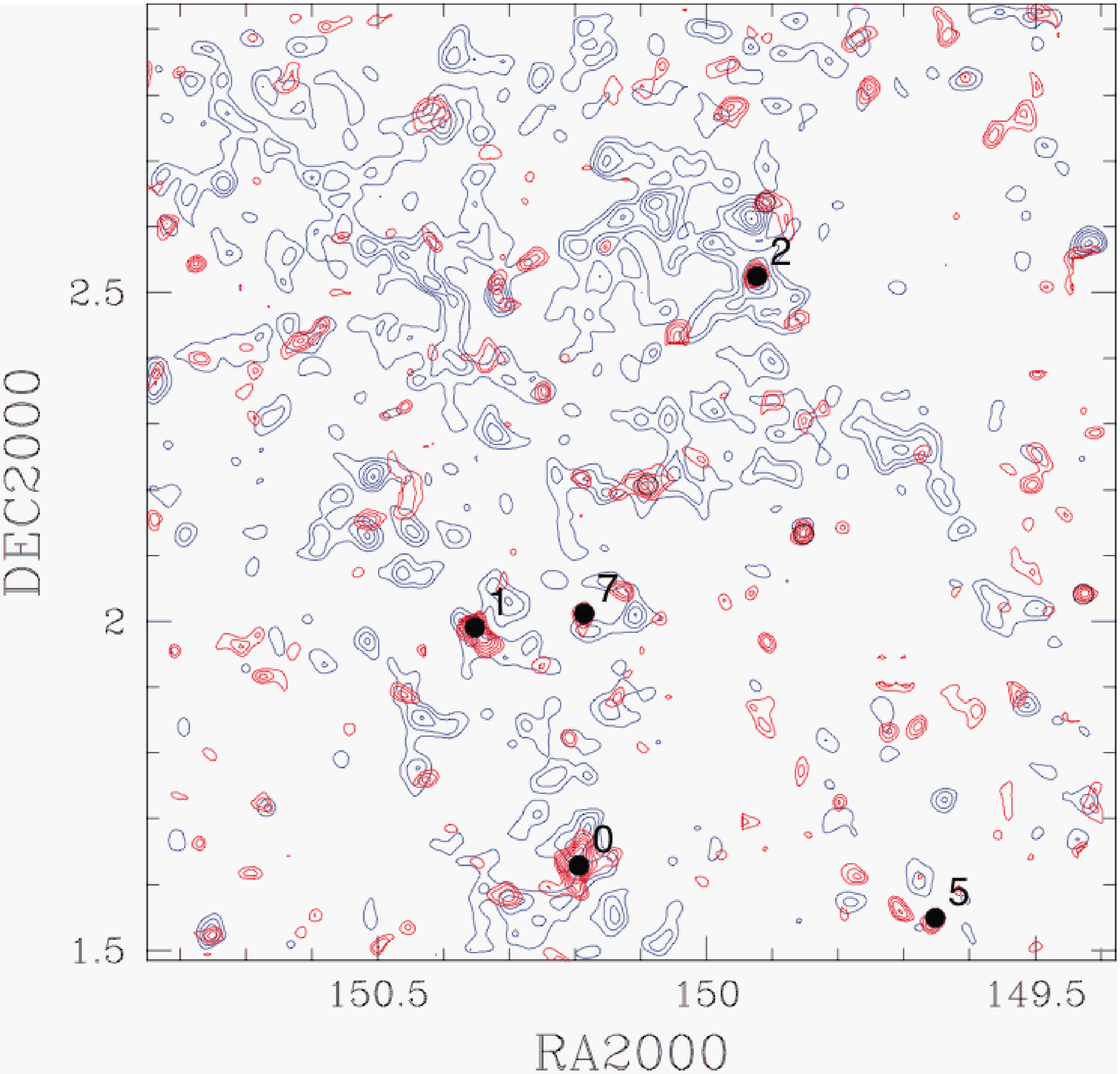}}}}
\figcaption{
COSMOS field $\kappa$-S/N map and surface density of galaxies.
\label{fig:cosmos_halomap}}
\end{center}
\end{figure*}

\begin{figure*}
\begin{center}
\centerline{{\vbox{\epsfysize=10.264cm\epsfbox{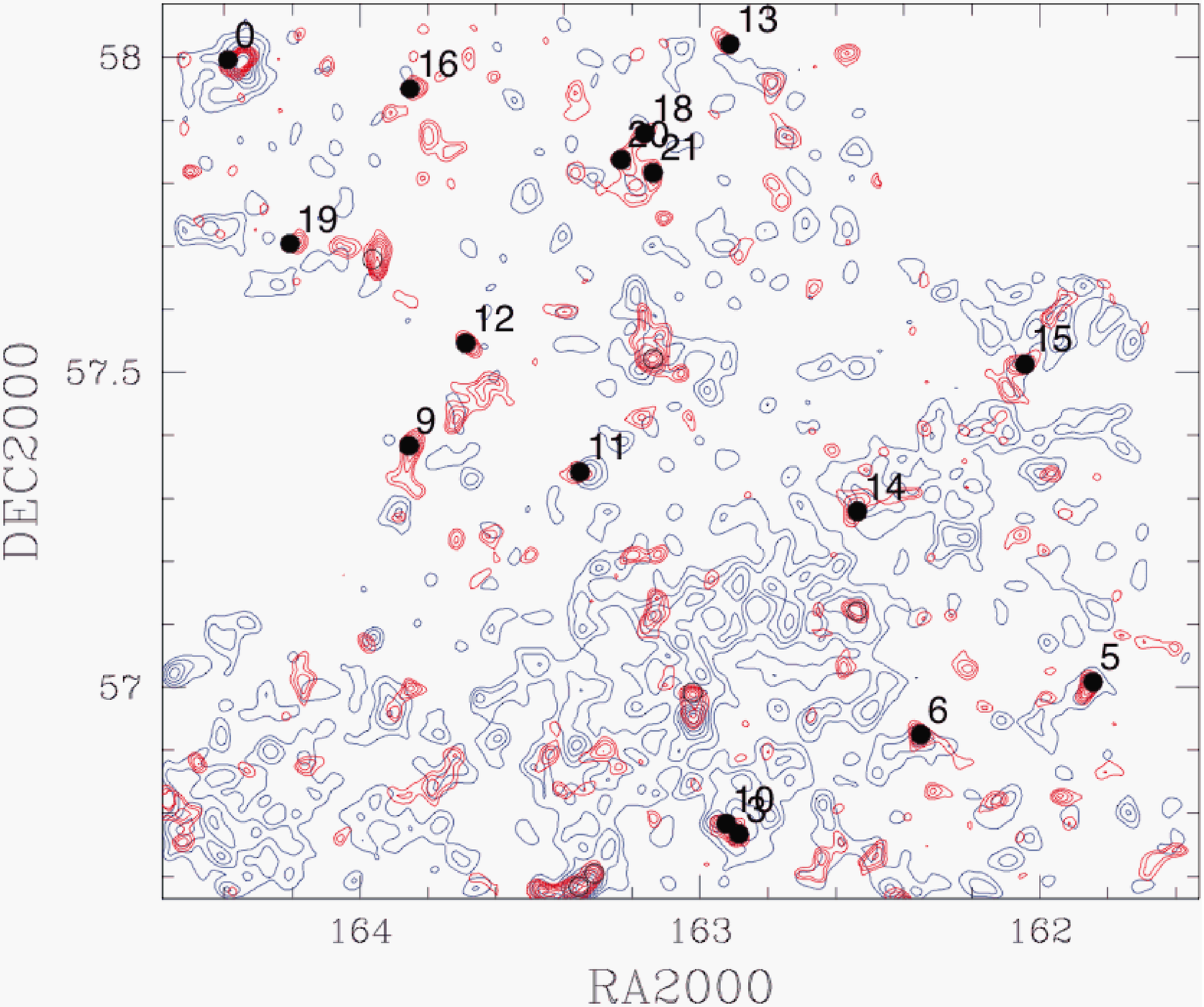}}}}
\figcaption{
Lockman field $\kappa$-S/N map and surface density of galaxies.
\label{fig:lockman_halomap}}
\end{center}
\end{figure*}

\begin{figure*}
\begin{center}
\centerline{{\vbox{\epsfysize=9.458cm\epsfbox{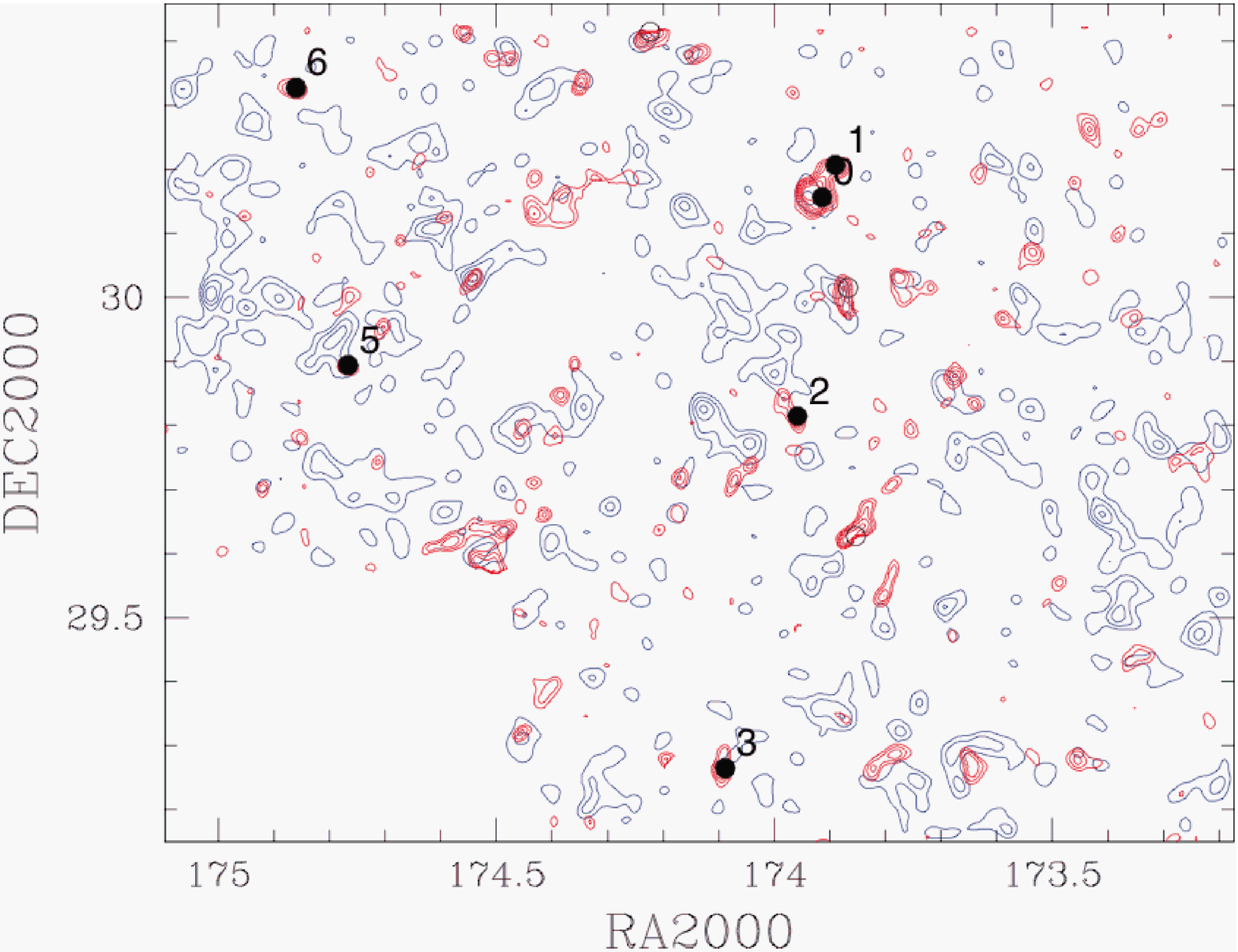}}}}
\figcaption{
GD140 field $\kappa$-S/N map and surface density of galaxies.
\label{fig:gd140_halomap}}
\end{center}
\end{figure*}

\begin{figure*}
\begin{center}
\centerline{{\vbox{\epsfysize=9.466cm\epsfbox{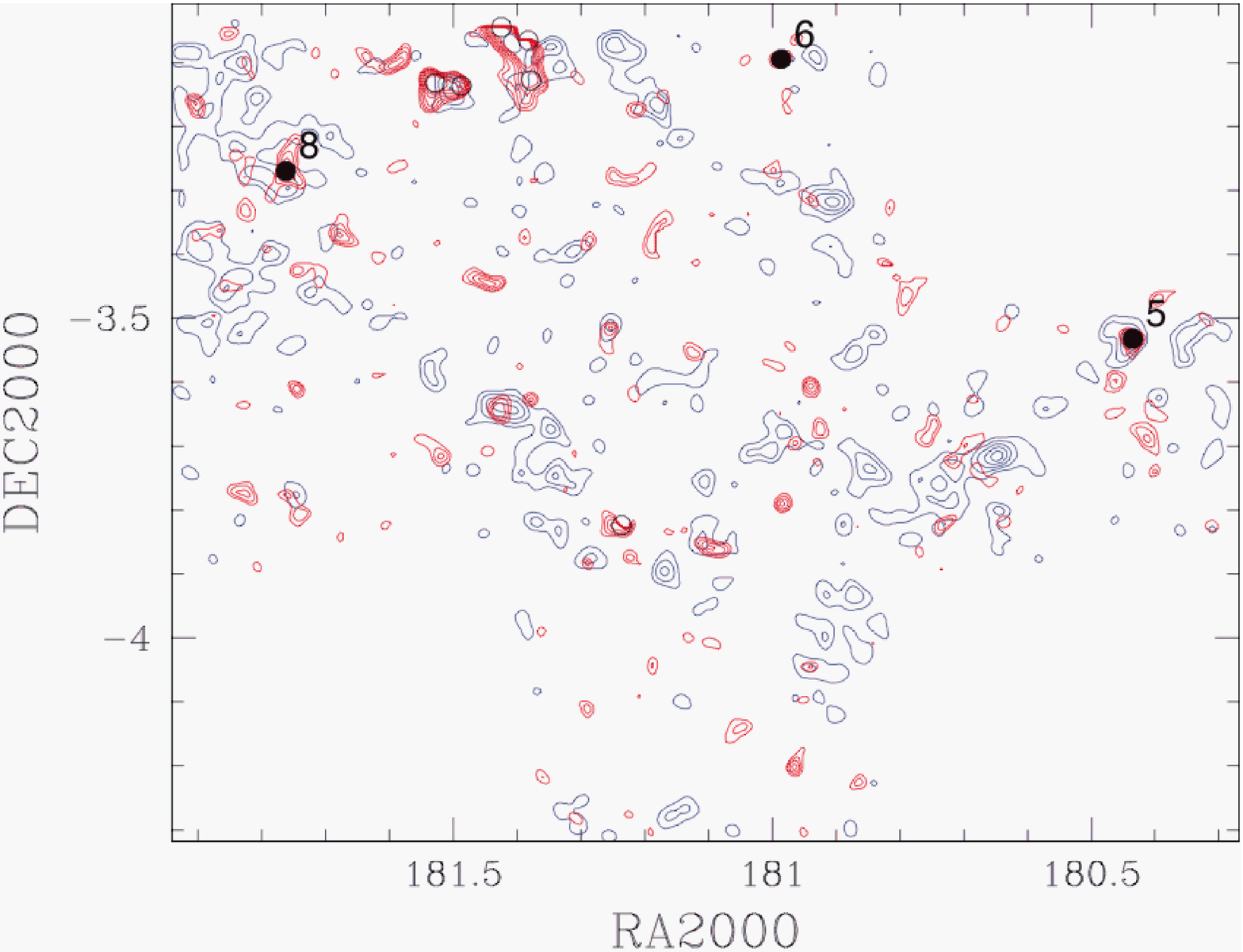}}}}
\figcaption{
PG1159-035 field $\kappa$-S/N map and surface density of galaxies.
\label{fig:pg1159-035_halomap}}
\end{center}
\end{figure*}

\begin{figure*}
\begin{center}
\centerline{{\vbox{\epsfysize=9.462cm\epsfbox{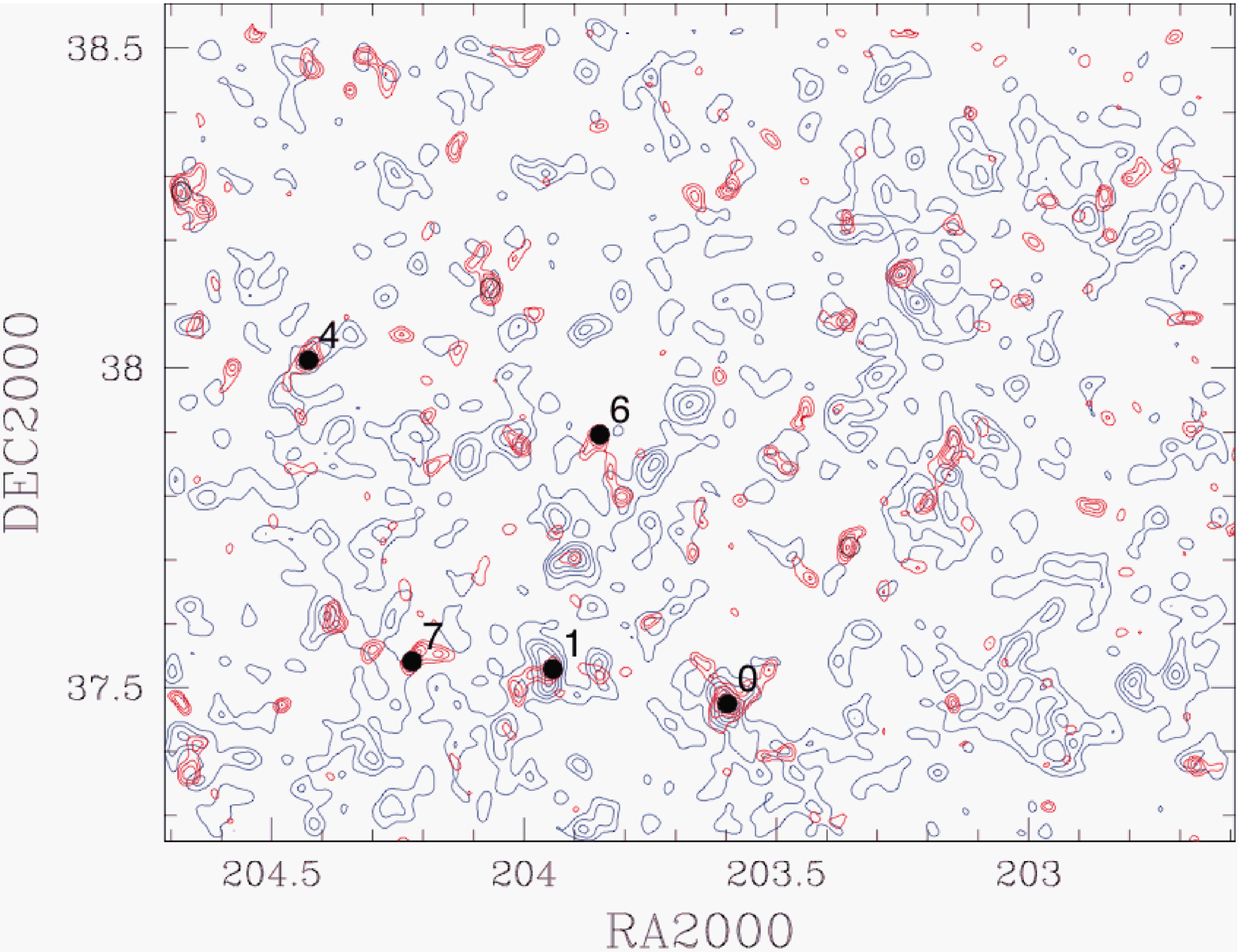}}}}
\figcaption{
13 hr field $\kappa$-S/N map and surface density of galaxies.
\label{fig:deep_survey_halomap}}
\end{center}
\end{figure*}

\begin{figure*}
\begin{center}
\centerline{{\vbox{\epsfysize=9.465cm\epsfbox{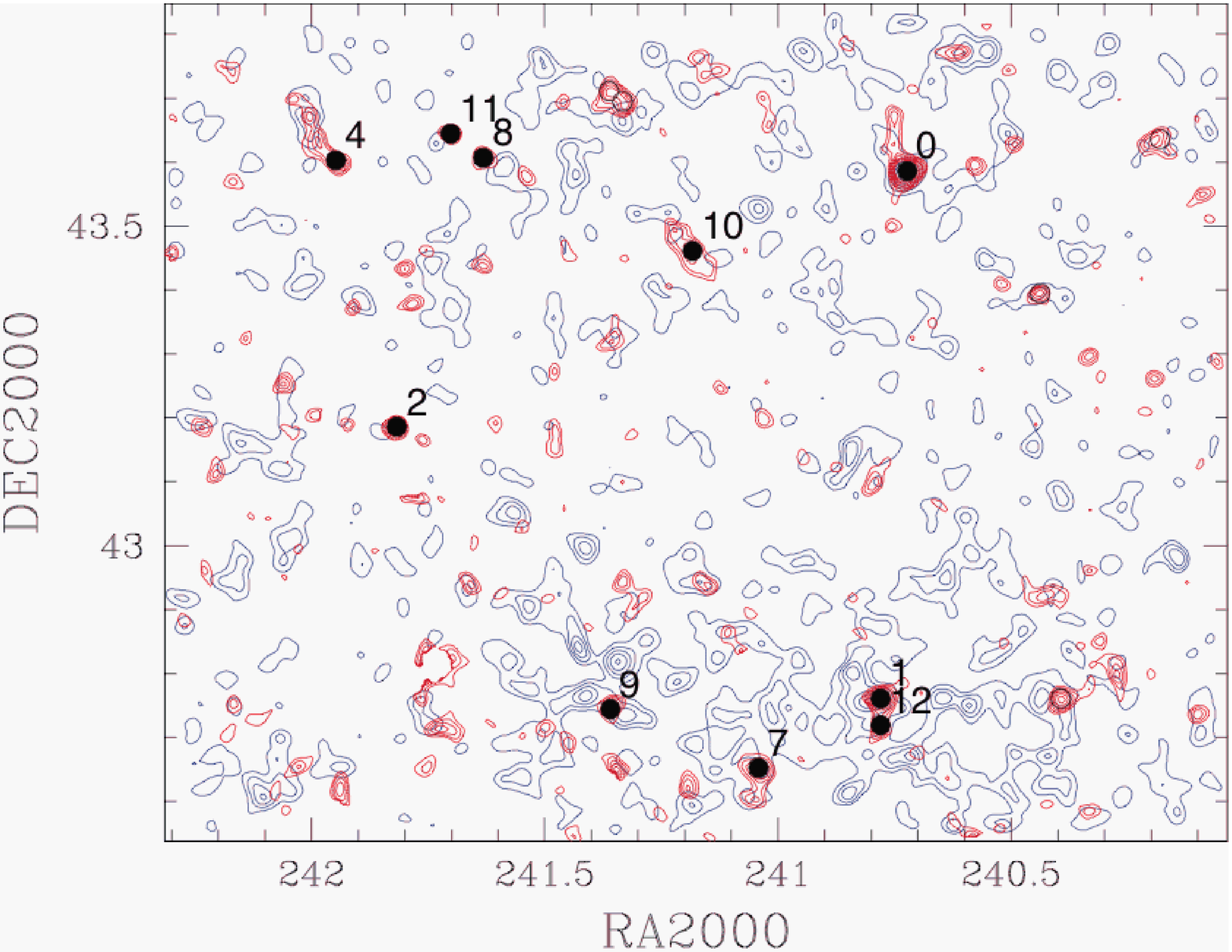}}}}
\figcaption{
GTO 2 square degree field $\kappa$-S/N map and surface density of galaxies.
\label{fig:gto_halomap}}
\end{center}
\end{figure*}

\begin{figure*}
\begin{center}
\centerline{{\vbox{\epsfysize=9.466cm\epsfbox{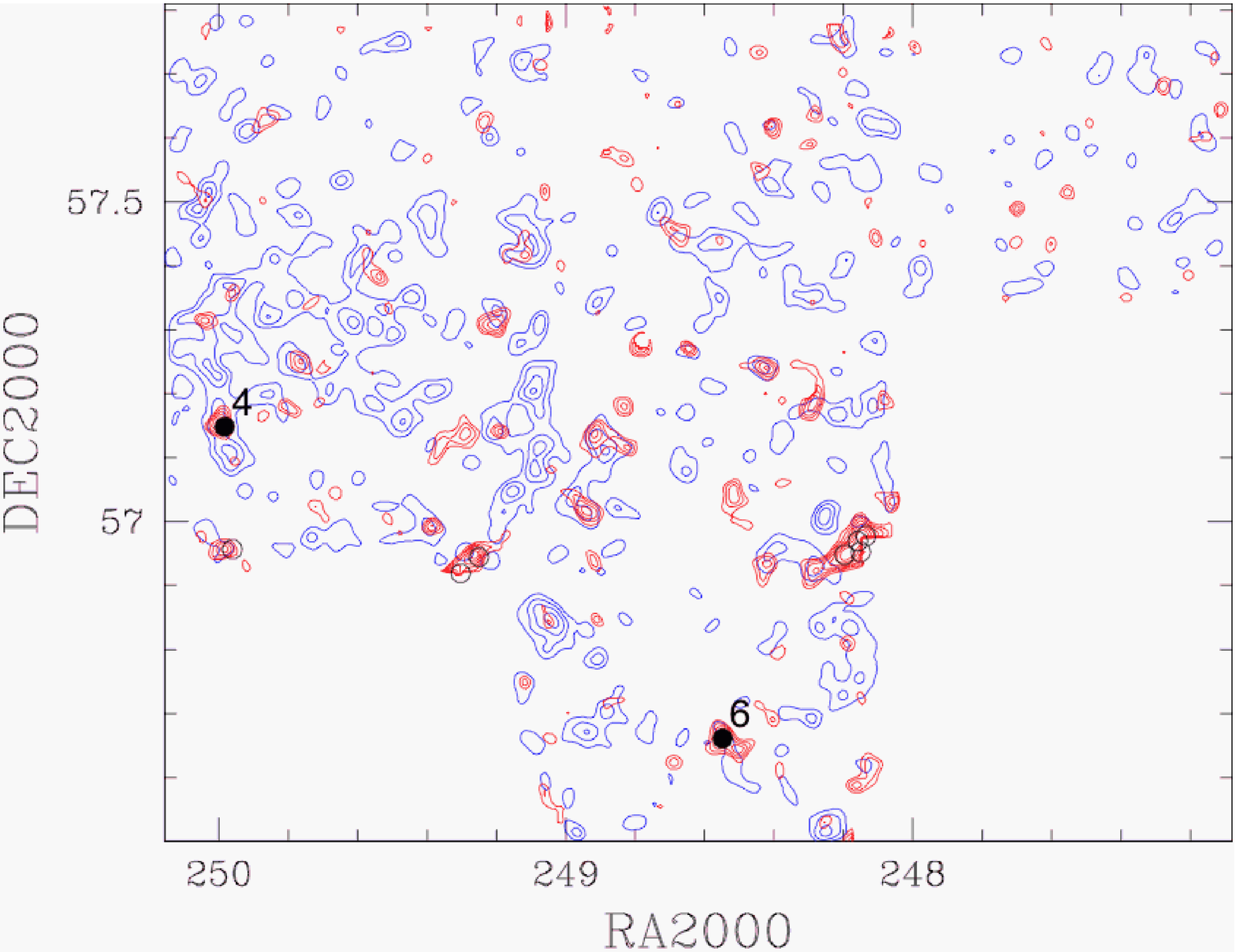}}}}
\figcaption{
CM DRA field $\kappa$-S/N map and surface density of galaxies.
\label{fig:cm_dra_halomap}}
\end{center}
\end{figure*}

\begin{figure*}
\begin{center}
\centerline{{\vbox{\epsfysize=4.326cm\epsfbox{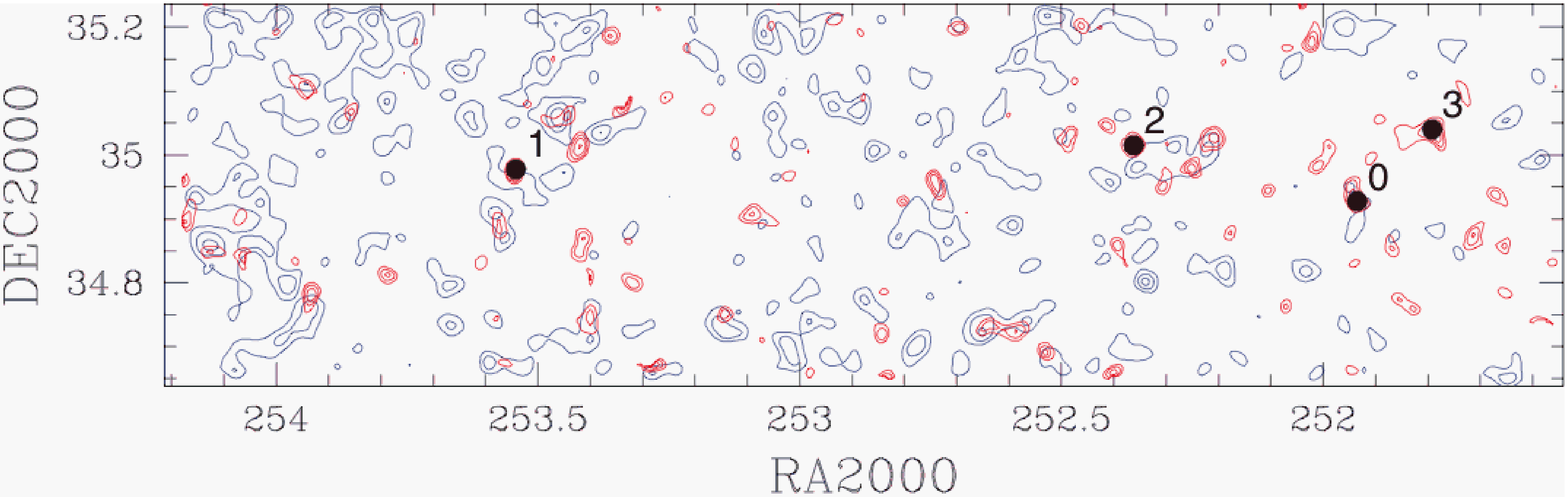}}}}
\figcaption{
DEEP16 field $\kappa$-S/N map and surface density of galaxies.
\label{fig:deep16_halomap}}
\end{center}
\end{figure*}

\begin{figure*}
\begin{center}
\centerline{{\vbox{\epsfysize=4.145cm\epsfbox{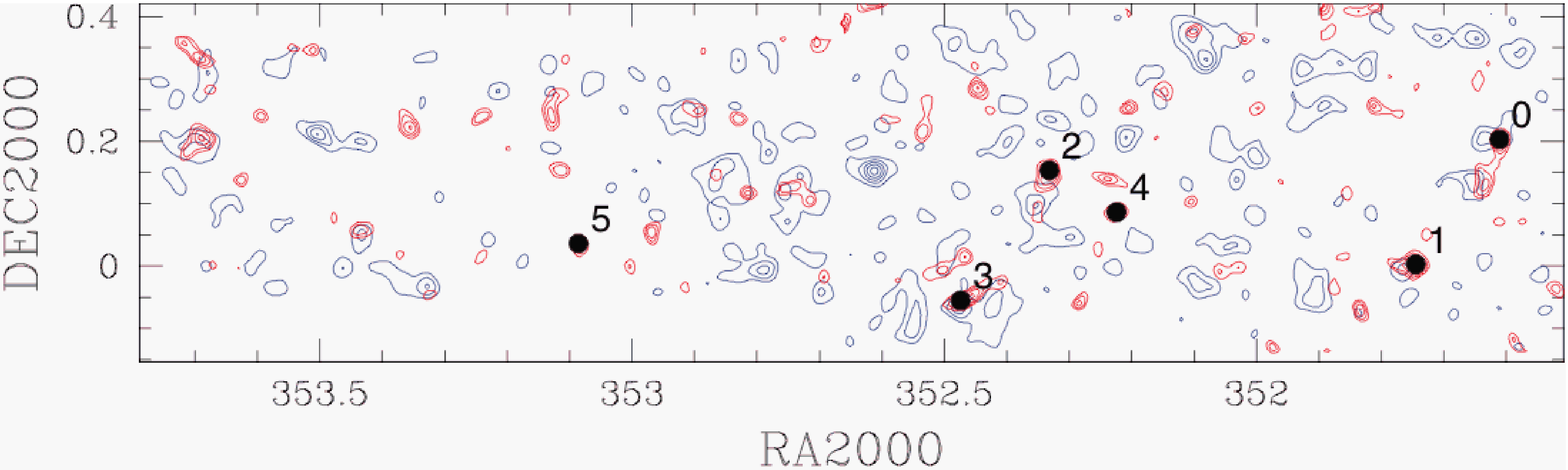}}}}
\figcaption{
DEEP23 field $\kappa$-S/N map and surface density of galaxies.
\label{fig:deep23_halomap}}
\end{center}
\end{figure*}

Figure~\ref{fig:saclay_halomap} shows the results of a halo search
in one of our fields: the XMM-Wide field.
The red contour shows the $\kappa$ S/N map where
the threshold and increment are set at 2 and 0.5, respectively. The
blue contour shows the surface number density of moderately bright
galaxies ($21 < R_c < 23$). Local peaks are searched on the
$\kappa$ S/N map and their positions are marked as open and filled
circles. Figure~\ref{fig:deep02_halomap} to \ref{fig:deep23_halomap}
show the $\kappa$ S/N maps for the remainder of our survey fields.

While visually inspecting galaxy concentrations around the detected halos, 
we noticed that less concentrated halos tend to occur preferentially near 
bright stars and field boundaries. Since regions near bright stars are 
masked (section \ref{sec:galaxycatalogs}) conceivably the discontinuity 
in the faint background galaxy distribution could cause spurious 
peaks. To avoid this, we reject halos occurring within 
a 4 arcmin radius of bright stars ($b_{USNO-A}$ $<$ 11) and within 2.3 
arcmin of the field boundary. These restrictions reduce the survey area by 
23 \% to what we will refer to as the {\it secure survey area} (16.72 deg$^2$).
Halos found within the secure area are termed the {\it secure sample}.
It is certainly possible that a significant fraction of halos lying in 
the non-secure area are genuine clusters. We will discuss this further
in a later paper concerned with their spectroscopic follow-up (Green et
al, in preparation). 

Detailed inspection of the halo candidates and the spectroscopic follow-up
discussed below revealed our completeness is high to a limiting
signal to noise in the convergence map of 3.69. Table~\ref{tab:halocat},
\ref{tab:halocat2} lists the secure sample with S/N $>$ 3.69. In this 
table, N$_g$ represents the number of moderately bright ($R_c < 22$) 
galaxies within 2 arcmin, indicative of the galaxy concentration.  

Concerning the optimum threshold for the significance, decreasing it
will increase the halo sample but likely introduce more false
detections. The optimum value should be set based on the
spectroscopically-observed true/false rate. Investigating the rate is
a major goal of our study.  In this work, the least significant
spectroscopically-identified halos have S/N = 3.69 (XMM-Wide n=23). We
adopted the threshold of 3.69 for this work so that all
the spectroscopically followed-up samples are included in the table.

We adopted significance maps for the selection of candidates
rather than kappa maps. This is because we would like to minimize
contamination by the false peaks. However, the effective kappa
threshold varies over the field, so we may encounter a
``completeness'' problem; i.e. halos that have high kappa value are
lost from the list. We investigated such omissions in the
XMM-Wide field (Figure~\ref{fig:saclay_halomap}), and only one halo is
found in this category, with $S/N < 3.69$ \& $\kappa > \kappa_{thres}$ 
where the $\kappa_{thres}$ is calculated as
$\kappa_{thres} = 3.69 \times Noise_{global}$.
The ``$Noise_{global}$'' is estimated globally over the entire
XMM-Wide field kappa map, and is 0.018 here. This halo is lost because
the local noise is as high as 0.022.

In practice, it will be very hard to generate completely uniform data
sets over the entire field of a survey because weather and seeing
conditions will vary. We will have to optimize the $\kappa_{thres}$ on
a field by field basis based on the data quality of each
field. Therefore, our strategy is the following: at first, we collect
reliable cluster samples based on the significance, and then users of
the catalog can set their own kappa threshold or mass threshold to
carry out their studies. This work represents the results of the first
step above.  We list the kappa values in Table~\ref{tab:halocat},
\ref{tab:halocat2} for reference.

\begin{table*}
\tabcaption{List of shear selected halos}
\label{tab:halocat}
\begin{footnotesize}
\begin{center}
\hspace*{-2.5em}
\begin{tabular}{llllllllllll}
\tableline\tableline\noalign{\smallskip}
Field & n & ID & RA & DEC & $\kappa$S/N & $\kappa$ & N$_g$\tablenotemark{a} & FOCAS \tablenotemark{b} & Known\tablenotemark{c} & NEDG\tablenotemark{d} & Note \\
\tableline\tableline\noalign{\smallskip}
DEEP02 &  00 & - &  37.32 &   0.63 & 4.44 & 0.099 &  16 & - & - & 1.35  &  \\
 &  01 & - &  37.87 &   0.51 & 4.39 & 0.083 & 24 & - & - & - &  \\
 &  02 & - &  37.38 &   0.41 & 4.25 & 0.120 & 19 & - & - & 0.73  &  \\
 &  04 & - &  37.15 &   0.73 & 4.11 & 0.086 & 25 & - & - & 0.10  &  \\
 &  05 & - &  37.31 &   0.44 & 4.07 & 0.131 & 20 & - & - & 1.03  &  \\
 &  06 & - &  37.72 &   0.69 & 3.97 & 0.081 & 18 & - & - & 0.86  &  \\
 &  07 & - &  37.86 &   0.57 & 3.95 & 0.074 & 18 & - & - & 0.92  &  \\
 &  08 & SL J0228.4+0030 &  37.12 &   0.51 & 3.93 & 0.085 &  49 & - & 0.46(P)  & - & VGCF 46 \\
 &  09 & SL J0228.2+0033 &  37.07 &   0.55 & 3.84 & 0.102 & 33 & - & 0.50(P) & - & {\tiny SDSS CE J037.099808+00.540769 }\\
\tableline\noalign{\smallskip} 
SXDS &  00 & - &  34.29 &  -5.59 & 5.33 & 0.057 &  26 & - & - & - &  \\
 &  01 & - &  34.38 &  -4.86 & 4.14 &  0.084 & 24 & - & - & - &  \\
 &  02 & - &  34.61 &  -4.41 & 3.96 &  0.059 & 16 & - & - & - &  \\
 &  03 & - &  34.74 &  -4.70 & 3.90 &  0.044 & 19 & - & - & - &  \\
 &  04 & - &  34.96 &  -5.12 & 3.86 &  0.069 & 21 & - & - & - &  \\
 &  05 & - &  34.41 &  -4.50 & 3.77 &  0.056 & 26 & - & - & - &  \\
\tableline\noalign{\smallskip} 
XMM-Wide &  00 & SL J0221.7-0345 &  35.44 &  -3.77 & 8.15 & 0.156  &  72 & - & 0.43  & - & XLSSC 006 \\
 &  01 & SL J0225.7-0312 &  36.43 &  -3.21 & 5.72 &  0.108 & 41 & 0.14  & - & - &  LRIS z = 0.14 \\
 &  02 & SL J0224.4-0449 &  36.10 &  -4.82 & 5.06 & 0.074 &  40 & 0.49  & - & - &  \\
 &  04 & - &  35.34 &  -3.50 & 4.91 & 0.082 &  21 & - & - & - &  \\
 &  08 & SL J0222.3-0446 &  35.48 &  -3.80 & 4.33 & 0.081 & 29 & - & - & - &  LRIS z = 0.41 \\
 &  10 & -  &  36.25 &  -4.25 & 4.20 & 0.062 &  23 & - & - & - & \\
 &  12 & SL J0224.5-0414 &  36.13 &  -4.24 & 4.06 &  0.057 & 70 & 0.26  & - & - & LRIS z = 0.26 \\
 &  15 & SL J0225.3-0441 &  36.34 &  -4.70 & 3.94 &  0.091 & 34 & 0.26  & - & - &  \\
 &  16 & SL J0228.1-0450 &  37.03 &  -4.84 & 3.94 &  0.072 & 31 & 0.29  & - & - &  \\
 &  17 & SL J0226.5-0401 &  36.63 &  -4.02 & 3.90 &  0.079 & 37 & - & 0.34  & - & XLSSC 014 \\
 &  19 & SL J0227.7-0450 &  36.94 &  -4.85 & 3.81 &  0.064 & 43 & - & 0.29  & - & Pierre et al. (2006) \\
 &  20 & - &  35.98 &  -3.77 & 3.81 &  0.048 & 20 & - & - & - &  \\
 &  21 & SL J0228.4-0425 &  37.12 &  -4.43 & 3.80 &  0.055 & 49 & - & 0.43  & - & XLSSC 012 \\
 &  22 & SL J0225.4-0414 &  36.36 &  -4.25 & 3.72 &  0.073 & 43 & 0.14  & - & - &  \\
 &  23 & SL J0222.8-0416 &  35.71 &  -4.27 & 3.69 &  0.049 & 52 & 0.43,0.19,0.23  & - & - &  \\
\tableline\noalign{\smallskip} 
Lynx &  00 & - & 131.91 &  44.80 & 5.84 & 0.121 &  20 & - & - & - &  \\
 &  01 & - & 132.59 &  44.07 & 5.01 & 0.083 &  43 & - & - & - &  \\
 &  03 & - & 131.83 &  44.86 & 4.57 & 0.139 & 23 & - & - & - &  \\
 &  05 & - & 131.77 &  44.85 & 4.37 & 0.110 &  13 & - & - & - &  \\
 &  07 & - & 132.69 &  44.95 & 4.15 & 0.105 & 31 & - & - & - &  \\
 &  08 & SL J0850.5+4512 & 132.64 &  45.20 & 4.02 & 0.085 &  53 & 0.19  & 0.24(P) & - & {\tiny NSC J085029+451141,LRIS$z=0.19$} \\
 &  09 & - & 131.47 &  44.96 & 4.02 & 0.076 & 26 & - & - & - &  \\
 &  10 & - & 133.02 &  44.14 & 4.00 & 0.108 & 23 & - & - & - &  \\
 &  12 & - & 132.37 &  44.38 & 3.90 & 0.081 & 36 & - & - & - &  \\
 &  13 & - & 132.41 &  44.37 & 3.90 & 0.072 & 31 & - & - & - &  Part of n=12\\
 &  14 & - & 132.54 &  44.07 & 3.86 & 0.066 & 48 & - & - & - &  \\
 &  15 & - & 132.81 &  44.35 & 3.77 & 0.077 & 39 & - & - & - &  \\
 &  16 & - & 132.31 &  44.30 & 3.75 & 0.072 & 44 & - & - & - &  \\
 &  17 & - & 131.40 &  44.94 & 3.74 & 0.084 & 25 & - & - & 0.15  &  \\
\tableline\noalign{\smallskip} 
COSMOS &  00 & SL J1000.7+0137 & 150.19 &   1.63 & 6.11 & 0.113 &  64 & 0.22  & 0.20(P)  & - & NSC J100047+013912 \\
 &  01 & SL J1001.4+0159 & 150.35 &   1.99 & 5.64 & 0.098 &  32 & - & 0.85(P)& - & \cite{finoguenovetal06}\\
 &  02 & SJ J0959.6+0231 & 149.92 &   2.52 & 4.74 & 0.067 &  83 & - & 0.73(P) & - & \cite{finoguenovetal06} \\
 &  05 & - & 149.65 &   1.55 & 3.92 & 0.078 & 47 & - & - & - &  \\
 &  07 & - & 150.19 &   2.01 & 3.88 & 0.070 & 36 & - & - & - &  \\
\tableline\noalign{\smallskip} 
\end{tabular}
\tablenotetext{a}{Number of moderately bright ($R_C < 22$) galaxies
around the halo within 2 arcmin}
\tablenotetext{b}{redshift obtained by FOCAS MOS}
\tablenotetext{c}{cluster redshift found in literatures (mainly from NED). ``P'' stands for photometric redshift.}
\tablenotetext{d}{redshift estimated from grouping of galaxies whose redshifts are listed on NED.}
\end{center}
\end{footnotesize}
\end{table*}

\begin{table*}
\tabcaption{List of shear selected halos (continued from Table~\ref{tab:halocat}) }
\label{tab:halocat2}
\begin{footnotesize}
\begin{center}
\hspace*{-2.5em}
\begin{tabular}{llllllllllll}
\tableline\tableline\noalign{\smallskip}
Field & n & ID & RA & DEC & $\kappa$S/N & $\kappa$ & N$_g$\tablenotemark{a} & FOCAS \tablenotemark{b} & Known\tablenotemark{c} & NEDG\tablenotemark{d} & Note \\ \tableline\tableline\noalign{\smallskip} 
Lockman &  00 & SL J1057.5+5759 & 164.39 &  58.00 & 6.28 & 0.109 &  68 & 0.60  & - & - &  \\
 &  03 & SL J1051.5+5646 & 162.88 &  56.77 & 4.97 & 0.082 &  31 & 0.33, 0.35 & - & - &  \\
 &  05 & SL J1047.3+5700 & 161.84 &  57.01 & 4.56 & 0.103 &  56 & 0.30, 0.24 & - & - &  \\
 &  06 & SL J1049.4+5655 & 162.35 &  56.93 & 4.51 & 0.095 &  47 & 0.42 & - & - &  LRIS z = 0.31 \\
 &  09 & SL J1055.4+5723 & 163.86 &  57.38 & 4.22 & 0.086 & 20 & - & - & - & LRIS z = 0.38 \\
 &  10 & SL J1051.6+5647 & 162.92 &  56.78 & 4.20 & 0.068 & 49 & 0.33, 0.25 & - & 0.05  &  part of SL J1051.5+5646 \\
 &  11 & SL J1053.4+5720 & 163.35 &  57.34 & 4.08 & 0.064 & 50 & - & 0.34  & - & RX J1053.3+5719 \\
 &  12 & - & 163.69 &  57.55 & 4.07 & 0.068 & 26 & - & - & - &  \\
 &  13 & - & 162.91 &  58.02 & 4.04 & 0.089 &  21 & - & - & 0.08  &  \\
 &  14 & - & 162.54 &  57.28 & 3.93 & 0.070 & 38 & - & - & - &  \\
 &  15 & SL J1048.1+5730 & 162.04 &  57.51 & 3.89 & 0.071 &  35 & 0.32  & - & - &  \\
 &  16 & - & 163.85 &  57.95 & 3.83 & 0.072 & 24 & - & - & 0.02  &  \\
 &  18 & - & 163.16 &  57.88 & 3.77 &  0.069 & 26 & - & - & - &  \\
 &  19 & - & 164.21 &  57.70 & 3.77 &  0.070 & 22 & - & - & - &  \\
 &  20 & - & 163.23 &  57.84 & 3.73 &  0.102 & 29 & - & - & - &  \\
 &  21 & - & 163.14 &  57.82 & 3.72 &  0.111 & 19 & - & - & - &  \\
\tableline\noalign{\smallskip} 
GD140 &  00 & SL J1135.6+3009 & 173.91 &  30.16 & 4.98 & 0.126 &  35 & 0.21  & - & - &  \\
 &  01 & -  & 173.89 &  30.21 & 4.19 & 0.086 &  23 & - & - & - &  \\
 &  02 & - & 173.96 &  29.81 & 4.09 & 0.069 &  17 & - & - & - &  \\
 &  03 & SL J1136.3+2915 & 174.09 &  29.26 & 4.03 & 0.100 &  24 & - & - & - &  LRIS z = 0.20 \\
 &  05 & - & 174.77 &  29.89 & 3.86 & 0.111 & 26 & - & - & - &  \\
 &  06 & - & 174.86 &  30.33 & 3.83 & 0.080 & 21 & - & - & - &  \\
\tableline\noalign{\smallskip} 
PG1159-035  &  05 & SL J1201.7-0331 & 180.44 &  -3.53 & 4.42 & 0.077 &  49 & 0.52  & - & - & \\
 &  06 & - & 180.99 &  -3.09 & 3.90 & 0.119 & 21 & - & - & 0.09  &  \\
 &  08 & - & 181.76 &  -3.27 & 3.71 & 0.102 & 37 & - & - & - &  \\
\tableline\noalign{\smallskip} 
13 hr Field &  00 & SL J1334.3+3728 & 203.60 &  37.47 & 4.33 & 0.128 & 74 & 0.30  & 0.48(P) & - & NSCS J133424+372822 \\
 &  01 & SL J1335.7+3731 & 203.94 &  37.53 & 4.10 & 0.091 &  65 & 0.41  & - & - &  \\
 &  04 & SL J1337.7+3800 & 204.43 &  38.01 & 3.85 & 0.080 & 34 & 0.18  & - & - &  \\
 &  06 & - & 203.85 &  37.90 & 3.78 & 0.068 & 27 & - & - & - &  \\
 &  07 & - & 204.22 &  37.54 & 3.77 & 0.078 & 30 & - & - & - &  \\
\tableline\noalign{\smallskip} 
GTO 2deg$^2$ &  00 & SL J1602.8+4335 & 240.72 &  43.59 & 6.65 & 0.110 & 56 & 0.42  & - & - &  \\
 &  01 & SL J1603.1+4245 & 240.78 &  42.76 & 5.47 & 0.106 &  57 & - & -  & - & LRIS z = 0.18  \\
 &  02 & - & 241.82 &  43.19 & 5.17 & 0.133 &  21 & - & - & - &  \\
 &  04 & - & 241.95 &  43.60 & 4.49 & 0.093 & 17 & - & - & - &  \\
 &  07 & SL J1604.1+4239 & 241.04 &  42.65 & 4.19 & 0.083 &  39 & - & - & - & LRIS z = 0.30 \\
 &  08 & - & 241.63 &  43.61 & 4.16 & 0.093 &  25 & - & - & - &  \\
 &  09 & SL J1605.4+4244 & 241.36 &  42.74 & 4.09 & 0.064 &  36 & 0.22  & - & - & \\
 &  10 & - & 241.18 &  43.46 & 3.85 &  0.092 & 29 & - & - & - &  \\
 &  11 & - & 241.70 &  43.64 & 3.84 &  0.077 & 33 & - & - & - &  \\
 &  12 & SL J1603.1+4243 & 240.78 &  42.72 & 3.82 & 0.091 &  37 & - & - & - & Part of SL J1603.1+4245 \\
\tableline\noalign{\smallskip} 
CM DRA &  04 & SL J1639.9+5708 & 249.98 &  57.15 & 4.15 & 0.073 &  32 & 0.20  & - & - &  \\
 &  06 & SL J1634.1+5639 & 248.55 &  56.66 & 3.97 & 0.104 &  31 & 0.24  & - & - &  \\
\tableline\noalign{\smallskip} 
DEEP16 &  00 & SL J1647.7+3455 & 251.94 &  34.93 & 4.30 & 0.084 &  42 & 0.26  & - & - &  \\
 &  01 & - & 253.54 &  34.98 & 4.24 & 0.078 &  24 & - & - & - &  \\
 &  02 & - & 252.36 &  35.02 & 3.75 & 0.123 &  25 & - & - & - &  \\
 &  03 & - & 251.79 &  35.04 & 3.72 & 0.087 & 21 & - & - & - &  \\
\tableline\noalign{\smallskip} 
DEEP23 &  00 & SL J2326.4+0012 & 351.61 &   0.20 & 4.41 & 0.076 &  19 & 0.28  & - & - &  \\
 &  01 & - & 351.75 &   0.00 & 4.37 & 0.095 &  20 & - & - & - &  \\
 &  02 & - & 352.33 &   0.15 & 4.27 & 0.107 & 33 & - & - & 1.38  &  \\
 &  03 & - & 352.47 &  -0.06 & 4.11 & 0.079 &  34 & - & - & 0.07  &  \\
 &  04 & - & 352.22 &   0.09 & 3.87 & 0.084 & 21 & - & - & 1.37  &  \\
 &  05 & - & 353.09 &   0.04 & 3.70 & 0.066 & 22 & - & - & - &  \\
\tableline\noalign{\smallskip}
\end{tabular}
\end{center}
\end{footnotesize}
\end{table*}

\vspace{0.3cm}
\centerline{{\vbox{\epsfxsize=8.5cm\epsfbox{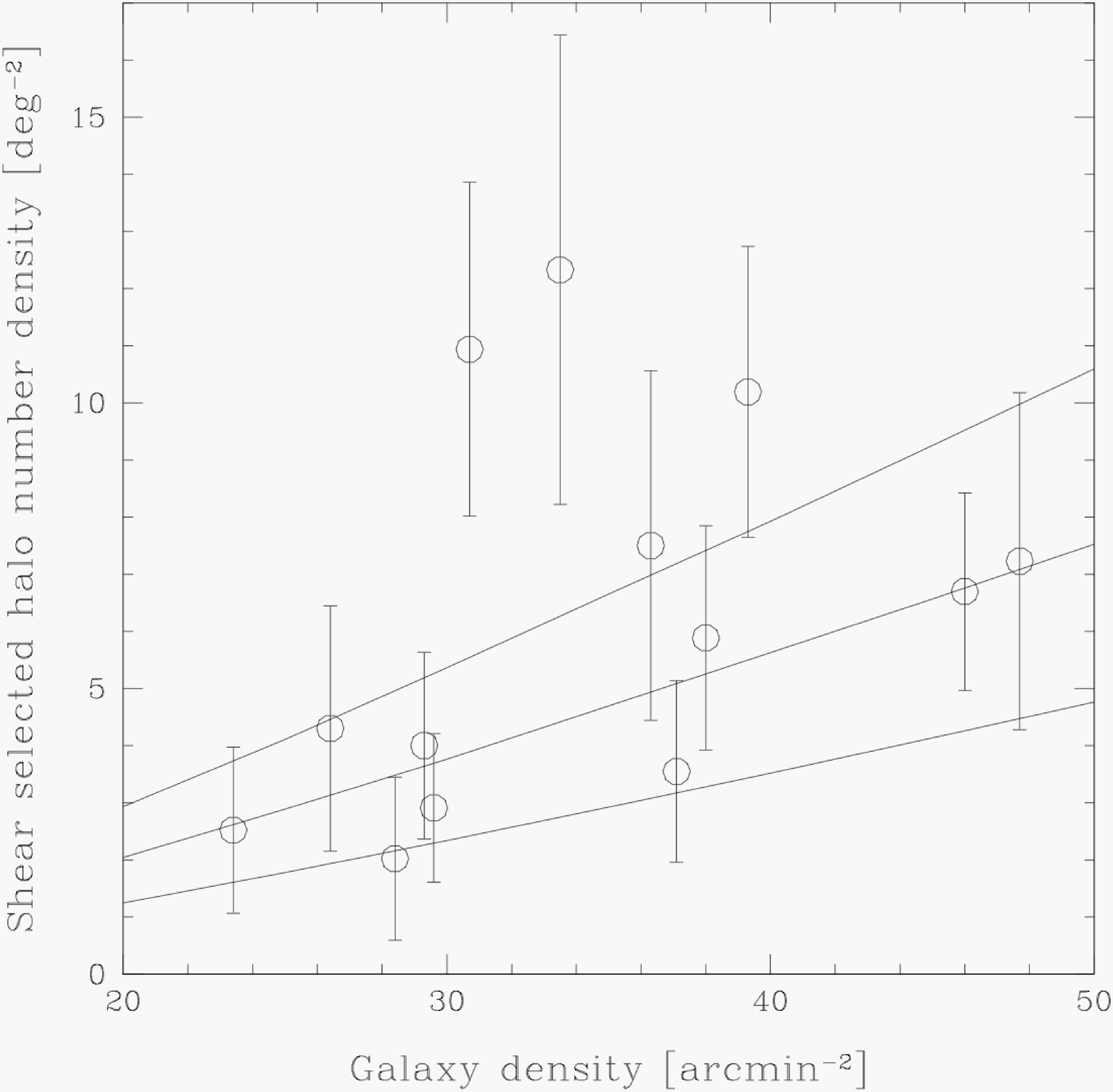}}}}
\figcaption{
Faint galaxy density ($\rho_{gal}$)  dependence of number density of
secured halos (S/N $>$ 3.69) in each field. The error bars are based
on $\sqrt{N}$ error estimates. Solid lines shows the predicted halo
number density. The galaxy distribution is assumed 
$
{dn \over { dz}} =
{\beta \over {z_\ast\Gamma[(1+\alpha)/\beta)]}}
\left( {z \over {z_\ast}} \right)^{\alpha}
\exp\left[ -\left( {z \over {z_\ast}} \right)^{\beta} \right]
$
where $\alpha=2.0$ and $\beta=1.5$ is adopted.
The mean redshift, $\langle z \rangle$ is related to $z_\ast$ by 
$\langle z \rangle 
= z_\ast \Gamma[(2+\alpha)/\beta] / \Gamma[(1+\alpha)/\beta]$ 
Three different $\langle z \rangle$ are assumed; 0.8, 1.0 and 1.2
(from bottom to the top solid line). \label{fig:seeingdens}} 
\vspace{0.3cm}

Here we compare the detected halo number density [deg$^{-2}$] with the
prediction of numerical simulations done by \cite{hamanaetal04},
who have attempted to reproduce a survey such as ours.
Figure~\ref{fig:seeingdens} shows the halo density of our thirteen
survey fields. The horizontal axis shows the density of faint galaxies
($\rho_{gal}$) used for weak lensing analysis. The error bars
are based only on $\sqrt{N}$ estimates and do not include the effects
of cosmic variance. Solid lines in Figure~\ref{fig:seeingdens} show
the prediction of the simulation where three different mean redshifts
for the background galaxies are assumed (from bottom to top: 0.8, 1.0
and 1.2).

Although the scatter is large, there is reasonable agreement between
the observations and the prediction. We also see a gradual
increase of the observed number density over the $\rho_{gal}$ range
sampled as expected. These comparisons validate our observational
procedures. We have, however, two outliers in
Figure~\ref{fig:seeingdens}. These could be caused by cosmic variance
or may arise from some other reasons. Further follow-up studies will
be important to clarify the issue.

\section{Cluster Identification and Verification}

Armed with the shear-selected halo catalog, we now discuss the tests
we have made to verify its reliability, using both spectroscopic observations
and comparisons with X-ray data in the fields where the overlap of
targets can be studied.  

\subsection{Spectroscopic Follow-up }

Because the lensing kernel (or window function) has a
relatively broad redshift range, the superposition of multiple low
mass halos with different redshifts could yield highly significant
weak lensing signals. Such `superposition halos' are unwelcome in a
mass-selected cluster catalog. In order to identify such superposition
halos we undertook a more comprehensive spectroscopic survey using a
multi-object spectrograph for selected candidates. This also
gives us the velocity dispersion of member galaxies, which is an
estimate of dynamical mass of clusters. By comparing dynamical mass
with weak lensing mass we will be able to discuss the dynamical state
of the clusters (Hamana et al. in preparation).

\begin{figure*}
\begin{center}
\includegraphics[height=8.5cm]{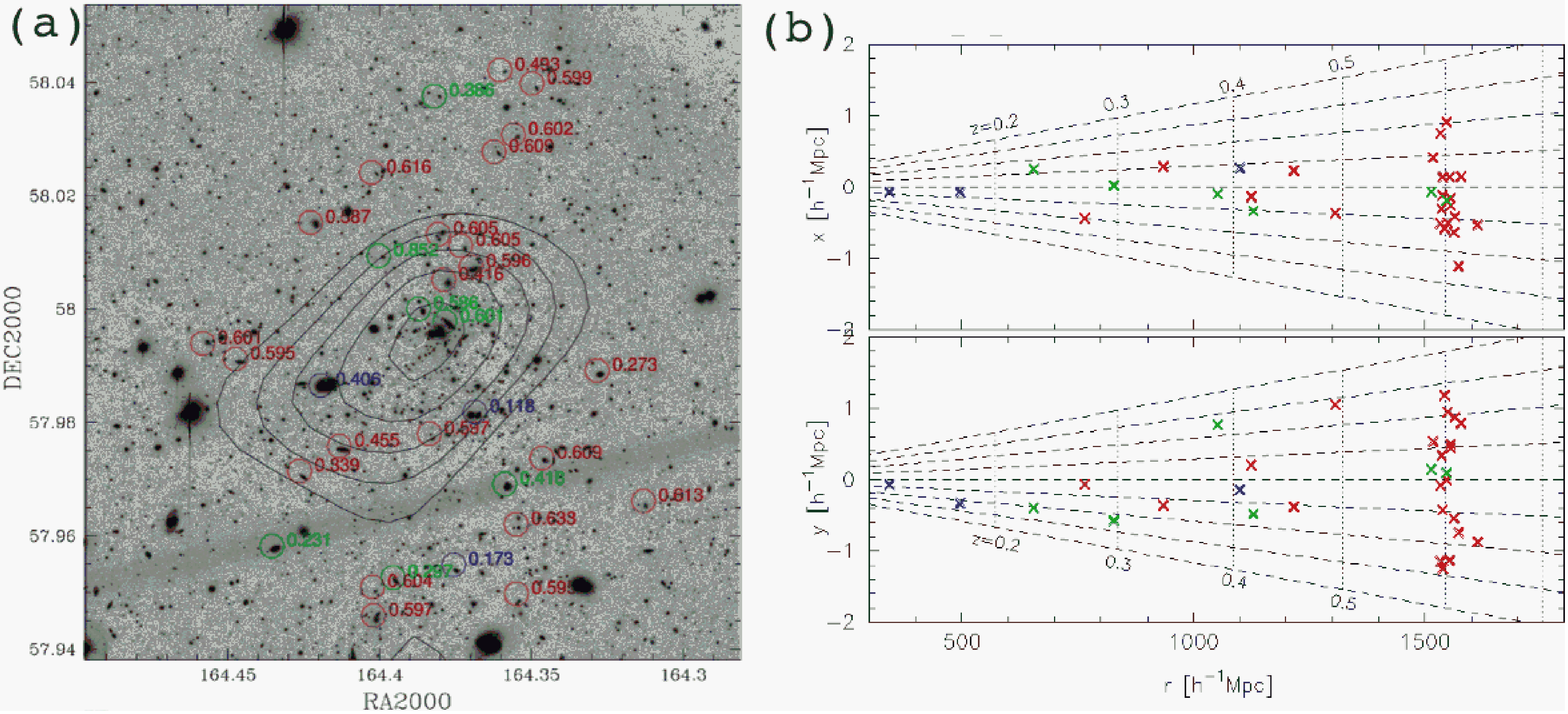}
\figcaption{
(a) $\kappa$ S/N contour map of the most significant halo in the
Lockman field (SL J1057.5+5759) superimposed on the $R_c$-band image
taken by Suprime-Cam. Contours start at a S/N = 2 2 with an interval of
1. Small circles show the positions of observed galaxies by FOCAS 
with the redshifts obtained.  (b) Redshift `cone diagram' of the
observed galaxies.
\label{fig:focasobs}}
\vspace{0.3cm}
\end{center}
\end{figure*}

We used the FOCAS spectrograph on Subaru whose multi-object mode
permits the simultaneous observation of 25 - 30 galaxies in the field
of a particular halo. We used the 150/mm grating and a SY47 order
sorting filter. This configuration spans the wavelength range
$4700-9400$\AA\ \citep{kashikawaetal02}. Target selection was based
primarily on apparent magnitude and the  color information was used
when available. The exposure time was 45 $\sim$ 70
minutes depending on the magnitude of the selected galaxies and the
observing conditions. The spectroscopic data was reduced using
standard IRAF procedures (Hamana et al. in
preparation). Figure~\ref{fig:focasobs} shows a typical FOCAS
observation where the target is identified as a $z$ = 0.6 cluster.

Since May 2004 we have observed 26 halos from the secure sample
with FOCAS. Higher priority was given to more significant halos,
except in the early stages of program (for example in the Lynx and
PG1159-035 fields) when the follow-up strategy was still being
evaluated. Each of the 26 halos has been reliably identified with a
cluster of galaxies. The redshift so determined is shown in the column
labeled ``FOCAS'' of Table~\ref{tab:halocat},\ref{tab:halocat2}.

In parallel with the FOCAS follow-up, long slit observations with
Keck/LRIS have been carried out to enlarge the sample of redshifts as
quickly as possible. We have to keep in mind that this method cannot
discriminate the projected samples, and some statistical consideration
is necessary in dealing with the data for further studies.
A complete discussion of that aspect of our program is discussed in a
separate paper (Green et al, in prep.). In the ``Note'' column of
Table~\ref{tab:halocat},\ref{tab:halocat2}, we list the preliminary
Keck/LRIS results as ``LRIS z = zvalue'' whose identification is
already robust (e.g. at least two galaxy redshifts agree)  including
one of the halos in the XMM-Wide field (SL J0222.3-0446, $z$=0.41)
where we discuss the reliability of our catalog (see section
\ref{sec:reliability}).

\subsection{Correlation with Published Data}

In addition to undertaking our own spectroscopic observations, we
also searched the NASA Extragalactic Database (NED) for clusters
and groups with published redshifts close to the position of our detected halos. 
We included recently-published cluster catalogs in the XMM-wide survey 
by \cite{valtchanovetal04}, \cite{willisetal05} and \cite{pierreetal06}
({\bf VWP} data hereafter). In making assignments, we considered
a cluster to be associated with a halo if the angular separation was less
than 2 arcmin and the 3 dimensional distance less than 1 $h^{-1}$ Mpc.
13 out of 100 halos can be identified in this way with
rich clusters in the literature and the published 
redshift is shown in the column labeled ``Known'' of 
Table~\ref{tab:halocat},\ref{tab:halocat2}. 
Three of these thirteen clusters were also observed 
by FOCAS in multi-object mode and we note that the mean FOCAS 
redshift is different from the published value for these systems 
($\delta z = 0.02 \sim 0.18$).  
These discrepancies can be explained by the fact that NED redshift of
these clusters are all photometric, which have larger
uncertainties. Therefore, we adopted the FOCAS redshifts for
these three cases. 
We searched the NED galaxy group catalog in the same manner, but no
group matched any of our halos.

Next, we searched for individual galaxies with published redshifts within 
3 arcmin of our halos. If suitable data was found, we grouped them together
using a ``friend of friend'' algorithm with a linking length of 
1 $h^{-1}$ Mpc. If we found a clustering of more than two members,
we calculated the average redshift and the mean astrometric position. 
Using the same proximity criterion above, we assigned a redshift
to 14 further halos. These are shown in the column labeled ``NEDG''.

In order to estimate the probability of a chance coincidence of the
matching procedures, we first randomized the position of the detected
halos in the field, and applied the same matching procedure described
above to the randomized halo catalog. This process suggests that
the probability of a chance coincidence is roughly 10 \% and 25 \% for 
the cluster and galaxy clustering searches, respectively. Clearly care 
should be taken in analyzing such data. 
The total number of reliably identified clusters are 41 (out of 100
in Table~\ref{tab:halocat},\ref{tab:halocat2}) where we do not include
those identified by ``NEDG'' galaxies because the
chance coincidence is not negligible. We assign ID labels only for
these reliable halos in the tables. 

\subsection{Reliability of the Shear-Selected Halo Catalog} 
\label{sec:reliability}
We now turn to the important question of the evaluating the reliability
of our shear-selected catalog. We will do this by examining both 
the success rate of our identifications and by comparing with X-ray samples 
obtained via the XMM-Wide field survey. VWP have confirmed
spectroscopic redshifts for 28 X-ray selected clusters in the 3.5
deg$^2$ survey field. Their locations are plotted as squares on
Figure~\ref{fig:saclay_halomap}. Our secure 
sample of 15 shear-selected halos with S/N $>$ 3.69 is plotted as
solid circles. 
Those we have spectroscopically confirmed using Subaru's FOCAS 
and/or Keck LRIS are marked by large open circles. We find that three
halos 
(1, 16 and 23) are confirmed as clusters by FOCAS and are not 
reported in VWP. 

Among our 15 shear-selected halo samples, only three (4, 10 and 20)
have yet to be identified. This identification success rate (80 \%) 
can be regarded as a lower limit given future observations may yet
locate an associated cluster. Although affected by small number
statistics, this minimum efficiency is already higher than expected 
by simulation
studies\citep{whiteetal02,hamanaetal04,hennawiandspergel05}. 

It is interesting to compare our identification success rate with
those found in the {\it GaBoDS} survey. \cite{maturietal06} found 14
significant halos in their 18 deg$^2$ area. Among them, 5 halos
turn out to be {\it known} clusters of galaxies, 2 seem to have
associated {\it light} concentrations but no spectroscopic confirmation, and 
the remaining {\it uncertain} 7 (50 \%) had no apparent counterpart in either 
optical nor X-ray data. \cite{schirmeretal06} undertook a search
on the same data adopting a different peak selection algorithm and
a lower detection threshold. They found 158 ``possible mass concentration'' 
on the 18 deg$^2$ field. If those halos are divided into the same classes
as above ({\it known, light, uncertain}), the ratio is almost the
same as \cite{maturietal06}.  Regardless, almost half of the {\it GaBoDS}
halos could be considered uncertain at this point. 

Meanwhile, we find 15 halos in a 2.24 deg$^2$ field (XMM-wide) and
show that 80 \% of have been already identified as clusters. 
It is too early to make any definite conclusion because their
spectroscopic is underway. However, our (tentatively) higher success
rate could be explained at least in part by the larger number density
of faint galaxies, $\rho_{gal}$, usable for the weak lensing
analysis owing to larger aperture and better average seeing.
In the case of {\it GaBoDS}, $\rho_{gal}$ spans from 6 to 28
arcmin$^{-2}$ depending on the field; the average value is 11
arcmin$^{-2}$. This is generally smaller than our
$\rho_{gal}$ shown in Table~\ref{tab:survey_field}. As a result one
can expect a lower angular resolution of the $\kappa$ map, reducing
the S/N ratio for a fixed smoothing scale and possibly an increased
degree of contamination in the resulting halo catalog.
Another possibility to explain the increased contamination is that
their sample consists of a combination of different sets of catalogs,
each of which is selected by different methods. This could decrease
the significance threshold effectively, and could introduce more false
peaks.

\subsection{Superposition of Multiple Clusters}
\cite{hamanaetal04} estimated, based on their simulation, that the 
halo superposition rate in survey such as ours should be roughly 3 \% -
a small but not negligible effect. A longslit spectroscopic survey, such
as that undertaken with LRIS (Green et al, in prep) might be poorly-suited
for locating such cases. However, the FOCAS multi-object survey reported
here should reliably find them. In fact, we have found
three apparent superposition halos  (SL J0222.8-0416, 
SL J1051.5+5456 and SL J1047.3+5700) out of 26 examined. 
The overlap rate is broadly consistent with expectation considering 
the small number so far sampled. 

\section{Conclusions and Future Prospects}

We have introduced a new Subaru imaging survey and described
techniques for locating and verifying shear-selected halos. Across
a search area of 16.72 deg$^2$ we have found 100 halos
whose $\kappa$ S/N exceeds 3.69. We have described the first
phase of a detailed follow-up campaign based on multi-object spectroscopy
of 26 halos using FOCAS on the Subaru telescope. A later paper in this
series (Green et al, in prep) will extend the spectroscopic survey to
the full sample using a longslit approach.

Detailed studies on one of our fields, the XMM-wide field, show that 
80\% of the shear selected 15 halos in the 2.2 deg$^2$ area can be
confirmed as genuine clusters of galaxies. 10 overlap with X-ray
detections and two are new systems confirmed spectroscopically. The
overall success rate and reliability of our sample provides convincing
proof that, with care, a weak lensing survey can provide a large 
sample of mass-selected halos. 

We compare our success rate
and the reliability of our catalog with that of the {\it GaBoDS} survey
and conclude a major advantage of our approach is the superior
imaging depth which leads to a high surface density of usable
galaxies. This suggests future, more ambitious, surveys for 
shear-selected halos will be more effective if undertaken with
large aperture telescopes.

It is interesting to use our results to estimate the requirements
for a future survey motivated by the need to constrain dark
energy. \cite{kolbetal06} discuss hypothetical missions which
would aim to analyze the redshift distribution $N(z)$ of 10,000
clusters. Figure~\ref{fig:seeingdens} shows that our 
survey technique typically 
finds 5 halos deg$^{-2}$. Because the field size of Suprime-Cam is 
$\sim$ 0.25 deg$^2$, it takes two hours to cover 1 deg$^2$ 
assuming the exposure time of 30 minutes of each field as adopted in
this study. A survey of 10,000 clusters would be prohibitive even
in terms of imaging alone ($\sim$ 500 clear nights) even before 
contemplating the follow-up spectroscopy. 
The Hyper Suprime-Cam (HSC) project \citep{hsc06} has been proposed 
to remedy this shortcoming. HSC is expected to have a field of view ten 
times larger than Suprime-Cam while maintaining the same image
quality. This new facility will makes the 2000 deg$^2$ scale imaging 
survey within a reasonable number of clear nights.

Spectroscopic follow-up might be enabled by
the proposed prime focus multi object optical spectrograph WFMOS 
(Wide Field Fiber Multi Object Spectrograph) whose field is 1.5 deg$^2$ 
in diameter. Typically we can expect $\simeq$10 shear-selected clusters 
in each spectroscopic field. With only 20 targets per halo, only a
small fraction of the several thousand fibers envisaged for WFMOS
need be allocated to the halo verification program.

\acknowledgments
We are very grateful to Subaru astronomers: Y. Oyama, K. Aoki
and T. Hattori for their dedicated supports of the FOCAS observing.
Numerical computations presented in this paper were carried out at 
the Astronomical Data Center (ADC) and at the Center for Computational
Astrophysics (CfCA) of the National Astronomical Observatory of
Japan. This work is supported in part by  Grant-in-Aid for Scientific
Research (Kaken-hi) of Japan Society for the Promotion of Science
(JSPS): Project number 15340065 (SM\&TH) and 17740116 (TH) .

\newpage

\appendix

\section{Probability to obtain spurious peaks due to insufficient
correction of the PSF anisotropy} \label{sec:imagequality}

\vspace{0.3cm}
The shear induced by massive cluster of galaxies is expected to be
7$\sim$ 10 \%. This is actually larger than raw ellipticities due to
PSF anisotropy (2 $\sim$ 4 \% refer to Fig.~\ref{fig:fwhmee}), but the 
difference is not so significant. Therefore, the correction of the
anisotropy is very important to make a precise kappa map. Here, we
examine the PSF anisotropy Suprime-Cam, and evaluate the effect of
imperfect correction in this work.

\vspace{0.3cm}
\centerline{{\vbox{\epsfxsize=17cm\epsfbox{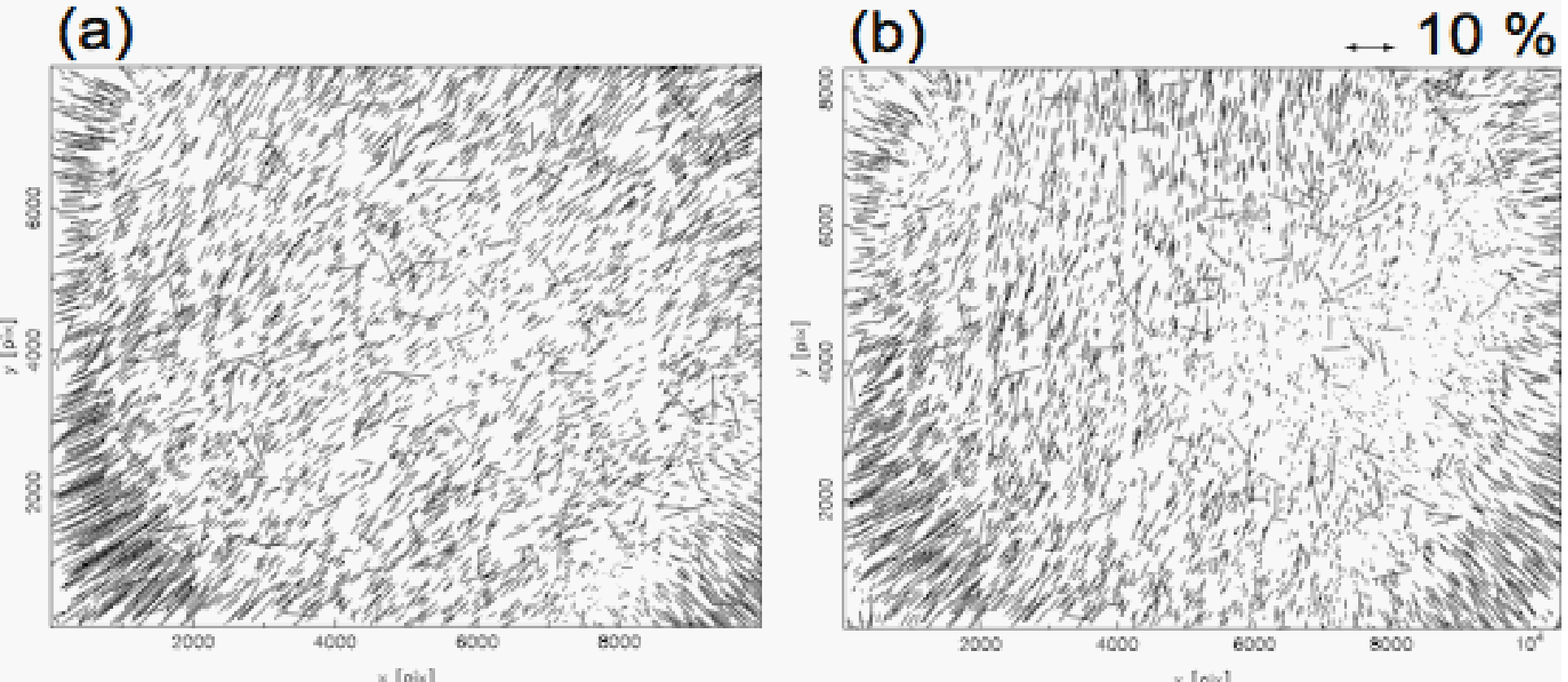}}}}
\figcaption{Two examples of ellipticities field. Orientation of each bar
shows the direction of the major axis, and the length scales as its
ellipticity. No subtraction of the offset is made. Arrow on left top
of the panel shows the size of 10 \% ellipticity. The seeing was 0.65
arcsec (FWHM).
\label{fig:starfield_e_examples}}
\vspace{0.3cm}

In order to characterize the anisotropy of the Suprime-Cam images, we
obtained sequences of i'-band short exposures (1 min) of dense stellar
fields with various telescope pointings over one night long. 
The seeing was mostly $\sim$ 0.65 arcsec (FWHM).
The PSF anisotropy is approximated as an elliptical, and the field
position dependence of the ellpticities are investigated. 
Two examples of such ellipticity fields are shown in
Fig.~\ref{fig:starfield_e_examples}. General tendency
that we note is that the fields can be represented as a
super-position of (1) the radial field at four corners (almost invariant)
and (2) almost unidirectional field (variable). 
The variable components is most likely due to the telescope shaking
whereas the invariant component can be explained by the optical
aberration of the corrector. Slight asymmetry of the radial component
is visible (i.e. ellipticities near the lower left corner is larger
than other corners), and this would be a sign of imperfect optical
alignment. 

\vspace{0.3cm}
\centerline{{\vbox{\epsfxsize=17cm\epsfbox{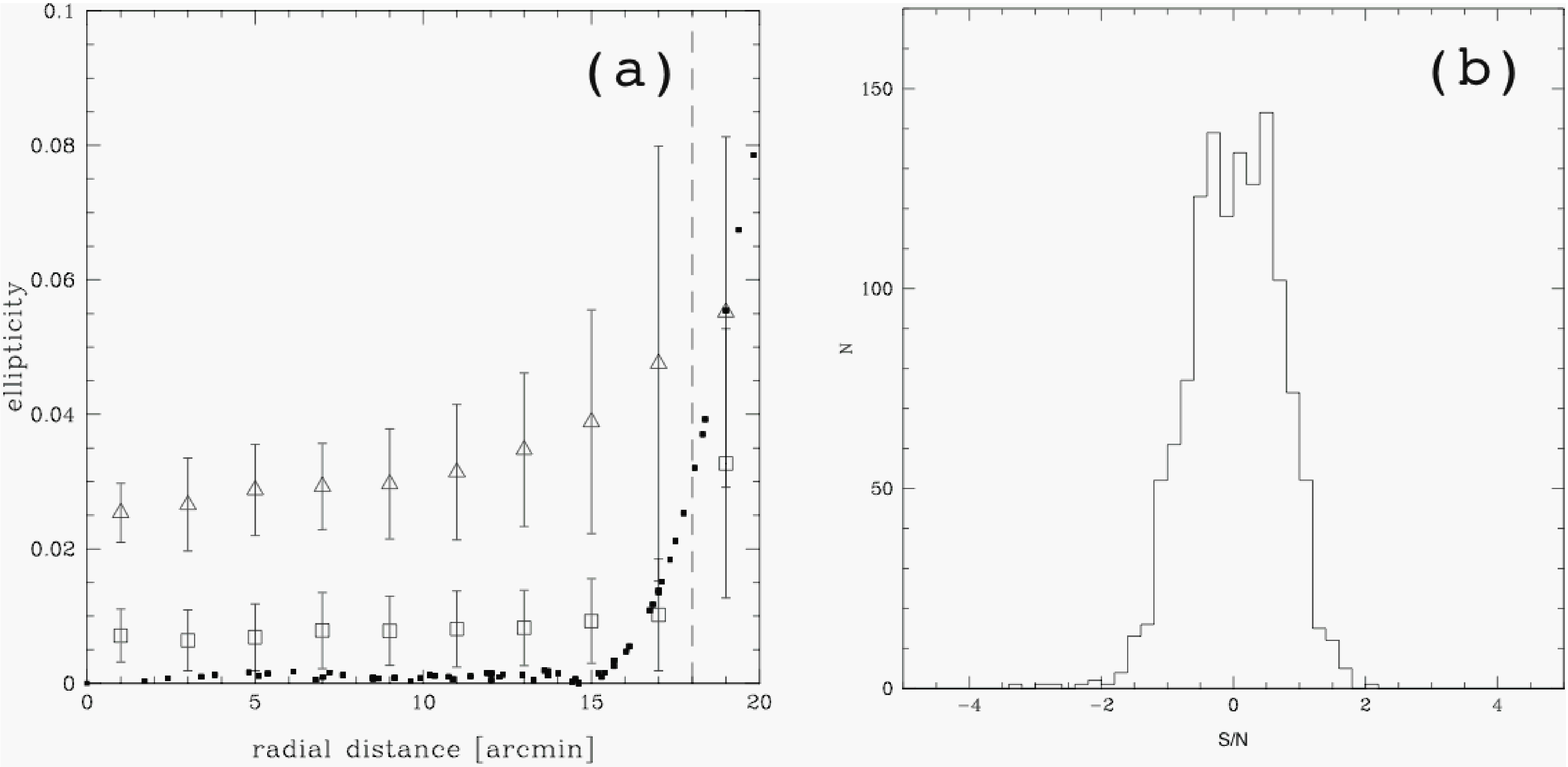}}}}
\figcaption{(a) Ellipticities of stars azimuthally averaged over the
  field of view (0.65 arcsec seeing i'-band 60 sec exposure).
Open circle and square show the results of ``before'' and ``after'' 
the anisotropy correction, respectively (see text).
Sold squares show expected ellipticities from perfectly aligned optics
under the seeing of 0.65 arcsec.
(b)Peak distribution function of kappa S/N map created by 
residual of anisotropy corrections. 
\label{fig:radreal}}
\vspace{0.3cm}

We also note that the discontinuity of ellipticities is not visible
across the boundary of the CCDs, and the ellipticities can be
represented as a single continuous function of field position. 
We adopt 5-th order polynomial function here. In order to simulate the
actual science field analysis, we randomly select controls stars and
``galaxy role'' stars from the star catalogs with an appropriate
density; 2 arcmin$^{-1}$ and 40 arcmin$^{-1}$, respectively. Using the
control stars, we obtain the best fit coefficients of the polynomial,
and the anisotropy of ``galaxy role'' stars is corrected using
Eqn.~\ref{Psmcorrection}. Because the intrinsic ellipticities of
``galaxy role'' stars is all zero, the residual ellipticities after
the correction is a estimate of incompleteness of the correction.
Fig~\ref{fig:radreal}(a) shows the azimuthally averaged
ellipticities of stars; before the correction, the typical 3 \% raw
ellipticity is seen. It might be interesting to note that the
ellipticities of 3 \%. can be induced by rms pointing error of 
merely 0.1 arcsec under the seeing of 0.65 arcsec. The ellipticities
is reduced down to about 0.75 \% after the correction. Beyond the
field angle of 18 arcmin, the correction does not work fine and we
decided to eliminate the field r $>$ 18 from lensing analysis; 
which, however,  results in only a few percent loss of FOV. 

We estimated a PSF anisotropy caused by the optical aberration using a
ray-tracing code, {\it zemax}. The calculated PSF is convolved with
0.65 arcsec FWHM gaussian, and the shape is evaluated by
elliptical. The result is shown in filled square in
Fig~\ref{fig:radreal}(a). Compared with this ideal case, the observed
ellipticities is large even after the correction. This shows a limit of
the adoption of single polynomial function as a representatives of the
PSF anisotropy, where the residual is still locally correlated
correlated and not completely random. We now want to evaluate the
impact of the incompleteness onto the $\kappa$ S/N map. 
The map is created based by the residual ellipticities after the
correction. We simulate the galaxy ellipticities using the following
``conversion'' formula \citep{hoekstra04},
\begin{equation}
\label{conversion}
e^{gal} = \frac{P_{sm}^{gal}}{P_{sm}^{*}}e^{*}, 
\end{equation}
which is essentially a sensitivity correction against the PSF
anisotropy. This is necessary because galaxies are larger and their
shape is more insensitive to the deformation compared with stars.
We evaluate $\frac{P_{sm}^{gal}}{P_{sm}^{*}}$ with
$\frac{\langle Tr(P_{sm}^{gal})\rangle}{\langle Tr(P_{sm}^{*})\rangle}$
where we adopt
$\langle Tr(P_{sm}^{gal}) \rangle = 0.1 $ which is a typical value
under the typical 0.7 arcsec seeing. Since the $\langle
Tr(P_{sm}^{*})\rangle \sim 0.2 $
here, the conversion factor is roughly $\frac{1}{2}$. We omit
$P_{\gamma}$ correction because it cancels out in this case when we
calculate the S/N. We created twelve such maps from independent
exposures, and co-added the peak distribution functions to obtain
Fig~\ref{fig:radreal}(b). It is obvious that the incompleteness of the
anisotropy correction is quite unlikely to  create any significant
(say S/N $>$ 3) fake peaks.


\begin{thebibliography}{999}

\bibitem[Allen et al.(2003)]{allenetal03}
Allen, S.W., Schmidt, R.W., Fabian, A.C., Ebeling, H. 2003 \mnras 342,
287

\bibitem[Allen et al.(2004)]{allenetal04}
Allen, S.W., Schmidt, R.W., Ebeling, H., Fabian, A.C., van Speybroeck,
L. 2004 \mnras, 353, 457

\bibitem[Becker et al.(2007)]{beckeretal07}
Becker, M.R. et al. 2007 submitted to ApJ (astro-ph/0704.3614)

\bibitem[Bernstein \& Jarvis (2002)]{bernsteinandjarvis02}
Bernstein, G. \& Jarvis, M. 2002, \aj, 123, 583

\bibitem[B\"ohringer et al.(2001)]{bohringeretal01}
B\"ohringer, H. et al. 2001 A\&A, 369, 826

\bibitem[Davis et al.(2003)]{davisetal03}
Davis, M., Faber, S.M., Newman, J. et al. 2003, SPIE, 4834, 161

\bibitem[Dietrich et al.(2007)]{dietrichetal07}
Dietrich, J.P., Erben, T., Lamer, G., Schneider, P., Schwope, A.,
Hartlap, J., Maturi, M (2007) A\&A in press (astro-ph/0705.3455)

\bibitem[Erben et al.(2001)]{erbenetal01}
Erben, T., van Waerbeke, L., Bertin, E., Mellier, \& Y. \& Schneider, P.
2001, A\&A, 366, 717

\bibitem[Finoguenov et al. (2006)]{finoguenovetal06}
Finoguenov, A., Guzzo, L., Hasinger, G., Scoville, N.Z. et al. (2006)
ApJS in press (astro-ph/0612360)

\bibitem[Gladders \& Yee(2000)]{gladdersetal00}
Gladders, M.D. \& Yee, H.K.C. 2000, \aj, 120, 2148

\bibitem[Goto et al.(2002)]{gotoetal02}
Goto, T. et al. 2002, \aj, 123, 1807

\bibitem[Hamana et al.(2003)]{hamanaetal03}
Hamana, T., Miyazaki, S. et al. 2003, \apj, 597, 98

\bibitem[Hamana et al.(2004)]{hamanaetal04}
Hamana, T., Takada, M., Yoshida, N. 2004, \mnras, 350, 893

\bibitem[Hennawi \& Spergel(2005)]{hennawiandspergel05}
Hennawi, J.F. \& Spergel, D.N. 2005 \apj, 624, 59

\bibitem[Henry (2000)]{henry00}
Henry, J.P. 2000, \apj, 534, 565

\bibitem[Hetterscheidt et al.(2005)]{hetterscheidtetal05}
Hetterscheidt, M., Erben, T., Schneider, P., Maoli, R., van Waerbeke,
L., Mellier, Y. 2005, A\&A, 442, 43. 

\bibitem[Heymans et al.(2006)]{heymansetal06}
Heymans, C. et al. 2006, \mnras, 368, 1323

\bibitem[Hoekstra et al.(1998)]{hoekstraetal98}
Hokekstra, H., Franx, M., Kuijken, K., Squires, G. 1998 \apj, 504. 636

\bibitem[Hoekstra (2004)]{hoekstra04}
Hoekstra, H. 2004, \mnras, 347, 1337

\bibitem[Ikebe et al.(2002)]{ikebeetal02}
Ikebe, Y., Reiprich, T.H., B\"ohringer, H., Tanaka, Y., Kitayama,
T. 2002 A\&A, 383, 773

\bibitem[Jenkins et al.(2001)]{jenkinsetal01}
Jenkins, A., Frenk, C. S., White, S. D. M., Colberg, J. M., Cole, S., 
Evrard, A. E., Couchman, H. M. P. \& Yoshida, N. 2001, \mnras,
324, 450

\bibitem[Kaiser \& Squires(1993)]{kaiserandsquires93} 
Kaiser, N. \& Squires, G. 1993, \apj, 404, 441

\bibitem[Kaiser et al.(1995)]{ksb95}
Kaiser, N., Squires, G. \& Broadhurst, T. 1995, \apj, 449, 460

\bibitem[Kaiser et al.(1999)]{kaiseretal99}
Kaiser, N., Wilson, G., Luppino, G., Dahle, H. 1999 submitted to PASP
 (astro-ph/9907229)

\bibitem[Kashikawa et al.(2002)]{kashikawaetal02}
Kashikawa, N. et al. 2002, \pasj, 54, 819

\bibitem[Kolb et al.(2006)]{kolbetal06}
Kolb et al. 2006, US Dark Energy Task Force Report
 
\bibitem[Landolt (1992)]{landolt92}
Landolt, A.U. 1992, \aj, 104, 340

\bibitem[Levine et al.(2002)]{levineetal02}
Levine, E.S., Schulz, A.E., White, M. 2002 \apj, 577, 569

\bibitem[Luppino \& Kaiser(1997)]{lk97}
Luppino, G.A. \& Kaiser, N. 1997, \apjl, 475, 20L

\bibitem[Majewski et al.(1994)]{majewskietal94}
Majewski, S.R., Kron, R.G., Koo, D.C., Bershady, M.A. 1994, \pasp,
106, 1258

\bibitem[Massey et al.(2007)]{massey07}
Massey, R.J, Rhodes, J., Ellis, R.S. et al 2007, Nature, 445, 286

\bibitem[Maturi et al.(2006)]{maturietal06}
Maturi, M., Schirmer, M., Meneghetti, M., Bartelmann, M., Moscardini,
L. 2006, A\&A in press (astro-ph/0607254)

\bibitem[Miyazaki et al.(2002a)]{miyazakietal02a} 
Miyazaki, S., Hamana, T., Shimasaku, Furusawa, H., Doi, M., Hamabe,
M., Imi, K., Kimura, M., Komiyama, Y., Nakata, F., Okada, N., Okamura,
S., Ouchi, M., Sekiguchi, M., Yagi, M., Yasuda, N. 2002a \apjl, 580, L97

\bibitem[Miyazaki et al.(2002b)]{miyazakietal02b} 
Miyazaki, S., Komiyama, Y., Okada, N., Imi, K., Yagi, M., Yasuda, N., 
Sekiguchi, M., Kimura, M, Doi, M., Hamabe, M., Nakata, F., Shimasaku,
K., Furusawa, H., Ouchi, M. \& Okamura, S. 2002b, \pasj, 54, 833

\bibitem[Miyazaki et al.(2006)]{hsc06}
Miyazaki, S., Komiyama, Y., Nakaya, H., Doi, Y., Furusawa, H.,
Gillingham, P., Kamata Y., Takeshi, K., Nariai, K. 2006, SPIE, 6269, 9

\bibitem[Pierre et al.(2004)]{pierre04}
Pierre, M. et al. 2004  J. Cosmol. Astropart. Phys, 09, 011

\bibitem[Pierre et al.(2006)]{pierreetal06}
Pierre, M., Pacaud, F. et al. 2006, \mnras, 372, 591

\bibitem[Postman et al.(1996)]{postmanetal96}
Postman, M., Lubin, L.M., Gunn, J.E., Oke, J.B., Hoessel, J.G.,
Schneider, D.P., Christensen, J.A. 1996, \aj, 111, 615

\bibitem[Press et al.(1993)]{pressetal93}
Press, W.H., Flannery, B.P., Teukolsky, S.A., Vetterling, W.T. 1993,
``Numerical Recipes in C'', ISBN-13: 9780521431088

\bibitem[Reiprich \& B\"ohringer(2002)]{reiprichetal02}
Reprich, T.H. \& B\"ohringer, H. 2002 \apj, 567, 716

\bibitem[Schirmer et al.(2006)]{schirmeretal06}
Schirmer, M., Erben, T., Hetterscheidt, M., Schneider, P. 2006, 
Submitted to A\&A (astro-ph/0607022)

\bibitem[Smail et al. (1997)]{smail97}
Smail, I., Ellis, R.S., Dressler, A. et al. 1997 \apj, 479, 70

\bibitem[Smith et al.(2005)]{smithetal05} 
Smith, G.P., Kneib, J., Smail, I., Mazzotta, P., Ebeling, H., Czoske,
O. 2005, \mnras, 359, 417

\bibitem[Valtchanov et al.(2004)]{valtchanovetal04}
Valtchanov, I. et al. 2004, A\&A, 423, 75

\bibitem[Wang et al.(2004)]{wangetal04}
Wang, S., Khoury, J., Haiman, Z. May, M. 2004 PhRvD, 70, 123008

\bibitem[Van Waerbeke et al.(2000)]{waerbeke00}
Van Waerbeke, L., Mellier, Y., Erben, T. et al. 2000, A\&A,
358, 30

\bibitem[White et al.(2002)]{whiteetal02}
White, M., van Waerbeke, L, Mackey, J. 2002, \apj, 575, 640

\bibitem[Willis et al.(2005)]{willisetal05}
Willis, J.P., Pacaud, F., Valtchanov, I. et al 2005, \mnras, 363, 675

\bibitem[Wittman et al.(2001)]{wittmanetal01}
Wittman, D., Tyson, J.A., Margoniner, V.E., Cohen, J.G., Dell'Antonio,
I.P. 2001 \apjl, 557, L89. 

\bibitem[Wittman et al.(2002)]{wittmanetal02}
Wittman, D, Tyson, J.A., Dell'Antonio, I.P. et al. 2002, Proc, SPIE, 4836, 73

\bibitem[Wittman et al.(2006)]{wittmanetal06}
Wittman, D, Dell'Antonio, I.P, Hughes, J.P. et al. 2006, \apj, 643, 128


\end{thebibliography}
\end{document}